\documentclass{article}
\usepackage{amsmath,amssymb,amsthm}
\usepackage{graphicx}
\usepackage{verbatim}

\usepackage{color}
\usepackage{makeidx}

 \makeatletter
 \@addtoreset{equation}{section}
 \makeatother

 \newtheorem{ittheorem}{Theorem}
 \newtheorem{itlemma}{Lemma}
 \newtheorem{itproposition}{Proposition}
 \newtheorem{itdefinition}{Definition}
 \newtheorem{ithyp}{Hypothesis}

 \newtheorem{itremark}{Remark}
 \newtheorem{itclaim}{Claim}
 \newtheorem{itcorollary}{\bf Corollary}

 \newenvironment{hyp}{\addtocounter{equation}{1}
 \begin{ithyp}}{\end{ithyp}}

 \newenvironment{theorem}{\addtocounter{equation}{1}
 \begin{ittheorem}}{\end{ittheorem}}

 \newenvironment{lemma}{\addtocounter{equation}{1}
 \begin{itlemma}}{\end{itlemma}}

 \newenvironment{proposition}{\addtocounter{equation}{1}
 \begin{itproposition}}{\end{itproposition}}

\def\bino{\text{Bin}}

\def \cH {{\cal H}}

\def \cP {{\cal P}}
\def \cQ {{\cal Q}}

\def \cA {{\cal A}}
\def \R {{\mathbb R}}

\def\PP{{\mathbb P}}
\def\EE{{\mathbb E}}

\def\bg{\overline{g}}
\def\oG{\overline{G}}
\def\of{\overline{f}}
\def\ofG{\overline{fG}}
\def \s {{\sigma}}

\def\exa{e^{-a}}
\def\zl{\{\,0,\dots,\ell\,\}}

\def\zu{\{\,0,1\,\}}
\def\ul{\{\,1,\dots,\ell\,\}}

\begin{document}

\title{Survival of the flattest in the quasispecies model}
\author{ Maxime Berger
\footnote{
\noindent
D\'epartement de math\'ematiques et applications, Ecole Normale Sup\'erieure,
CNRS, PSL Research University, 75005 Paris}
\footnote{
\noindent
Universit\'e de Bourgogne.
}
\hskip 70pt 
Rapha\"el Cerf 
\footnotemark[1]
\footnote{
Laboratoire de Math\'ematiques d'Orsay, Universit\'e Paris-Sud, CNRS, Universit\'e
Paris-Saclay, 91405 Orsay.}
}

\maketitle

\begin{abstract}
Viruses present an amazing genetic variability. 
An ensemble of infecting viruses, also called a viral quasispecies, is
a cloud of mutants centered around a specific genotype.
The simplest model of evolution,
	whose equilibrium state is described by the quasispecies equation,
is the Moran-Kingman model.
For the sharp peak landscape, we perform several exact computations
and we derive several exact formulas.
We obtain also an exact formula for 
the quasispecies distribution, involving a series and the mean fitness.
A very simple
formula for the mean Hamming distance
is derived, which is exact
and which does not require a specific asymptotic expansion
(like sending the length of the macromolecules to $\infty$
or the mutation probability to $0$).
With the help of these formulas,
we present an original proof for 
the well-known phenomenon of the error threshold.
We recover the limiting quasispecies distribution in the long chain regime.
We try also to extend these formulas to a general fitness landscape. We obtain an equation
involving the covariance of the fitness and the Hamming class number in the quasispecies distribution.
We go beyond the sharp peak landscape and we consider fitness landscapes having finitely many peaks and
a plateau-type landscape. We finally prove rigorously within this framework the possible occurrence of
the survival of the flattest, a phenomenon which has been previously discovered by
	Wilke, Wang, Ofria, Lenski and Adami \cite{Wixa}
	and which has been investigated in several works 
	\cite{CFJS,Frankl,FLAT,teja}.
\end{abstract}
\vfill\eject
\tableofcontents
\vfill\eject

\section{Introduction}
Some viruses are very hard to fight because of their genetic diversity: they mutate
constantly and this allows them to escape 
the attacks of the immune system of their host. 
At the same time, the genotypes of an infecting population of viruses are strongly 
correlated, they look like 
a cloud of mutants centered around a specific genotype.
This population structure, called a quasispecies, can be observed experimentally
for viruses ({\it in vivo} studies have been conducted for the HIV \cite{hiv} and the hepatitis C virus \cite{hepa}). 
Nowadays, deep sequencing techniques allow to collect data on the structure of viral quasispecies.
Yet the various biological mechanisms involved in the creation and the evolution of quasispecies are
far from being understood
(see \cite{dope} for a recent review or the book \cite{doschu} for a more comprehensive account).
The design of simple mathematical models of quasispecies
might help
to develop efficient antiviral therapies.
In principle, a quasispecies 
can occur in any biological population whose evolution is 
driven by mutation and selection.
In the next subsection,
we present a very simple model for the evolution of a population under mutation
and selection. We write then the equations describing the equilibrium of this model, thereby recovering directly
the quasispecies equations associated with the equilibrium of Eigen's model. All the results presented afterwards
concern the solutions
of these equations.
\subsection{The quasispecies equations}
We consider a population of haploid individuals evolving under the conjugate effects
of mutation and selection. 
We suppose that the population has reached an equilibrium
around a specific well adapted individual, 
called the wild type, and denoted by $w^*$.
Individuals reproduce, yet the reproduction mechanism is error-prone, and mutations
occur constantly. 
These mutations drive the genotypes away from~$w^*$.
Yet $w^*$ has a selective advantage because it reproduces faster.

We would like 
to characterize mathematically this kind of equilibrium.
We denote by $E$ the set of possible genotypes, which we assume to be finite. 
Generic elements of $E$ are denoted by
the letters $u,v$.
The Darwinian fitness of an individual having genotype $u$ is denoted by $f(u)$, it can be thought of
as its mean number of offspring. 
In addition, mutations occur in each reproduction cycle, an individual of type $u$ 
might appear as the result of mutations from offspring of other types.
Let us denote by $M(v,u)$ the probability that the offspring of an individual of
type $v$ is of type $u$. 
For simplicity, we will assume that
all the coefficients
of the mutation matrix $M$ are positive.
Of course, we have
$$\forall\, v\in E\qquad\sum_{u\in E}M(v,u)=1\,.$$

We introduce next the linear model, one of the simplest models
for the evolution of a population with selection and mutation.
We suppose that the successive generations do not overlap and
we denote by $N_n(u)$ the number of individuals 
of type $u$ in the generation $n$.
The linear model assumes that an individual
of type $v$ produces a number of offspring
proportional to its fitness $f(v)$,
and that a proportion $M(v,u)$ of the offspring
mutates and becomes of type $u$, thus $N_{n+1}(u)$ is given
by the formula
$$\forall\,u\in E\qquad
N_{n+1}(u)\,=\,\sum_{v\in E}N_n(v)f(v)M(v,u)\,.$$
The trouble with this formula is that the sum is not necessarily an integer.
To go around this problem, a natural way is to develop stochastic population models, in such a
way that the above formula describes the evolution of the mean number of individuals. The archetype
of this kind of models is the Galton-Watson branching process. If we introduce in addition a constraint
on the total size of the population, then we would consider the classical Wright-Fisher model.
Yet the randomness adds an extra layer of complexity and stochastic models are considerably harder to study. 
Another simpler possibility is to consider the proportions of each type of individuals in the population,
instead of their numbers, as 
Moran and Kingman did in the late seventies~\cite{Kingman,MO1}.
Let us
denote by $x_n(u)$ the proportion of individuals of type $u$ 
in the generation $n$.
The
model proposed by Moran is given by
\begin{equation*}
\forall\, u\in E\qquad
x_{n+1}(u)\,=\,\frac{
\sum_{v\in E}x_n(v)f(v)M(v,u)}{
\sum_{v\in E}x_n(v)f(v)}\,.
\end{equation*}
The denominator has been chosen to ensure that 
$\sum_{u\in E}x_{n+1}(u)=1$.  
In fact, this model consists in iterating a deterministic map on the function
encoding the proportions of each type present in the population. 
The possible equilibria of the model correspond to the fixed points of this map,
that is the solutions of 
the following equations:
\begin{equation}
\label{qs}
\forall\, u\in E\qquad
x(u)\,\sum_{v\in E}x(v)f(v)\,=\,
\,\sum_{v\in E}x(v)f(v)M(v,u)\,
\end{equation}
subject to the constraint
\begin{equation}
\label{cons}
\forall\, u\in E\qquad x(u)\geq 0\,,\qquad
\sum_{u\in E}x(u)\,=\,1\,.  
\end{equation}
We call these equations the quasispecies equations.
They characterize also
the equilibrium in the model originally developed by Eigen \cite{EI1}, which led to the quasispecies theory.
Some of our results are derived for the general quasispecies
equations~\eqref{qs}.
To go further, 
we will focus on a particular choice of the set of genotypes $E$ 
and of the mutation matrix $M$. 
Both for practical and historical reasons,
we make the same choice as 
Eigen 
and Schuster \cite{ECS1}, which leads to the 
sharp peak landscape.
\medskip

\noindent
{\bf Genotypes.}
We consider the different genotypes to be sequences of length $\ell\geq1$ over 
an alphabet $\cA$ of cardinality $\kappa\geq 1$.
Standard examples are $\cA=\{\,0,1\,\}$, $\kappa=2$,
$E=\lbrace\,0,1\,\rbrace^\ell$
for binary sequences or
$\cA=\lbrace\,A, C, G, T\,\rbrace$, $\kappa=4$, 
$E=\lbrace\,A,C,G,T\,\rbrace^\ell$ to model
DNA molecules.
The $i$-th component of a genotype $u$ is denoted by $u(i)$.
The Hamming distance $d$
counts the number of
differences between two sequences: 
$$\forall\,u,v\in E\qquad
d(u,v)\,=\,\text{card}\,\big\lbrace\,
1\leq i\leq \ell:
u(i)\neq v(i)
\,\big\rbrace\,.$$
When we wish to work with the simplest model, we use binary sequences.
When we seek to state a more general result, we use sequences over
a finite alphabet. 
\medskip

\noindent
{\bf Mutations.}
We suppose that mutations happen independently along the sequence
with probability $q\in\,]0,1[\,$, and that, 
upon mutation, a new letter
of the alphabet is chosen with equal probability.
For $u,v\in E$,
the mutation probability $M(u,v)$ is
given by
\begin{equation}
\label{mut}
M(u,v)\,=\,\Big(\frac{q}{\kappa-1}\Big)^{d(u,v)}(1-q)^{\ell-d(u,v)}\,.
\end{equation}
\noindent
%
The final missing ingredient to specify completely the model
is the fitness function. 
We will consider different types of fitness functions in the
following subsections.
For the initial models, for which the fitness function is quite simple,
we work with the genotype space $E=\cA^\ell$
where $\cA$ is a finite alphabet.
For the more complicated fitness landscapes considered at the end
which contain a kind of plateau,
we work with
the genotype space 
of binary sequences 
$E=\lbrace\,0,1\,\rbrace^\ell$.
\subsection{The sharp peak landscape}
\label{dspl}
Ideally, we would like to have explicit formulas for $x$ in terms of $f$ and $M$.
Unfortunately,
there is little hope of obtaining such explicit formulas in the general case. 
So we consider 
the simplest non neutral fitness function
which comes to mind,
the sharp peak landscape:
there is a privileged genotype,
$w^*$,
referred to as the wild type,
which has a higher fitness than the rest.
We work with the genotype space $E=\cA^\ell$
where $\cA$ is a finite alphabet of cardinality $\kappa$. 
Let $\sigma>1$ and let 
the fitness function $f_{\text{SP}}$ be given by
\begin{equation}
\label{sp}
\forall\,u\in E\qquad
f_{\text{SP}}(u)\,=\,\begin{cases}
\quad\sigma&\quad\text{if}\quad u=w^*\,,\\
\quad1&\quad\text{if}\quad u\neq w^*\,.
\end{cases}
\end{equation}
This is the fitness function that Eigen studied in detail in his article~\cite{EI1}.
Despite its apparent simplicity, this model leads to interesting mathematical questions
and it provides new insight into the genetic structure of a population.
%
Eigen 
and Schuster \cite{ECS1}
discussed the sharp peak landscape (which is fully presented in section~\ref{mutdyn})
with the help of approximation techniques
and in a specific asymptotic regime.
We present here a new approach, which is more elementary.
The sharp peak landscape is studied in detail
in section~\ref{prof}. We perform several exact computations
and we derive several exact formulas.
In particular, we obtain
a remarkable equation for the mean fitness $\lambda$ at equilibrium, which had been previously discovered by
Bratus, Novozhilov and Semenov,
see formulas~$5.5$ and $5.6$ in \cite{BSN1}. 
We state next a probabilistic version of this formula.
Let $S_\ell$ be a random variable with distribution the binomial law with parameters $\ell$ and $(\kappa-1)/\kappa$.
Denoting by $\EE$ the expectation,
the mean fitness $\lambda$ at equilibrium satisfies
\begin{equation}
\label{innlambda}
	\frac{1}{\sigma-1}
	\,=\, \EE\left(
	\frac{1}{
	\displaystyle
		\frac{\lambda}{
\rho^{S_\ell}}-1}
	\right)\,,
\end{equation}
where we have set
\begin{equation}
	\label{fdefr}
	\rho\,=\,1-\displaystyle \frac{\kappa q}{\kappa-1}\,.
\end{equation}
Bratus, Novozhilov and Semenov
exploited this equation to derive rigorous bounds on the mean fitness, but
this formula has wider applications.
Furthermore, 
we obtain also in subsection~\ref{exqs} an exact formula for 
the quasispecies distribution, involving a series and the mean fitness.
A very simple formula
for the mean Hamming distance 
is derived, which is exact
and which does not require a specific asymptotic expansion, like sending the length of the macromolecules to $\infty$
or the mutation probability to $0$
(recall that the Hamming distance
between two sequences
is the number of sites where they
differ).
The mean Hamming distance $\overline{\cQ}$
in the quasispecies distribution can be interpreted
as the mean number of ones in a genotype at equilibrium, 
and the formula is
\begin{equation}
\label{imeanHamm}
	\overline{\cQ} \,=\, 
\frac{\ell q\, \lambda} {\lambda-\rho}
\,,
\end{equation}
where $\lambda$ is the mean fitness of the population
(here we assume that the wild-type is the genotype $(0,\dots,0)$ and the Hamming distances are computed with respect
to the wild-type).

We stress that all the formulas obtained in
section~\ref{prof} are exact and hold for a
fixed value of the length $\ell$. 
In the next subsection and in section~\ref{tet},
we 
perform asymptotic expansions of these formulas
in the long chain regime. 
\subsection{The error threshold}
We shall
demonstrate that the quasispecies equations for the sharp peak landscape undergo a phenomenon similar to a phase transition. This is a classical and well-known result. 
However we adopt an original approach for the proof, indeed
we will show how it follows easily from 
equation~\eqref{innlambda}.
Furthermore,
with the help of the exact formulas derived in section~\ref{exqs},
we recover the limiting quasispecies distribution in the long chain regime in section~\ref{lqd}.

We consider the asymptotic regime, called the long chain regime, where
\begin{equation}
\label{iasym}
\ell\to+\infty\,,\qquad
q\to 0\,,\qquad
\ell q\to a\in\,[0,+\infty]\,.
\end{equation}
The parameter~$a$ is in fact
the asymptotic mean number of observed mutations per individual in each reproduction cycle.
We denote by $\lambda$ the mean fitness at equilibrium and by $x(w^*)$ the proportion
of the wild type $w^*$ in the population at equilibrium.
\begin{theorem}[Error threshold]
  \label{ierrt}
  For the sharp peak landscape,
we have the following dichotomy:

\noindent
$\bullet$ If $a>\ln \sigma$, then
$\lambda
\rightarrow 1$, whence $x(w^*)\rightarrow 0$.

\noindent
$\bullet$ If $a<\ln \sigma$, then
$\lambda
\rightarrow \sigma e^{-a}>1$, and
\begin{equation*}
x(w^*)\quad\to\quad\frac{\sigma\exa-1}{\sigma-1}\,.
\end{equation*}
\end{theorem}
\noindent
Thus the limit of the proportion $x(w^*)$ of the wild type present in the population is positive
only if $\sigma\exa>1$,
or equivalently, if asymptotically $q<\ell^{-1}\ln\sigma$.
The value
$$q^*\,=\,\frac{\ln\sigma}{\ell}$$
is the error threshold,
described by Eigen as the critical mutation rate above which
the wild type $w^*$ disappears from the population.
Whenever the occurrence of mutations is negligible compared to the
sequence length,
that is for $a=0$, we have $x(w^*)=1$ and
the only genotype present in the population is the wild type $w^*$.
In the presence of a small number of mutations, that is for $a>0$,
genetic diversity is constantly reintroduced in the population
and there is a positive proportion of the genotypes which differs from $w^*$. Yet most 
of the genotypes are very close to the wild type $w^*$.
In fact, the population looks like a cloud of mutants centered around
the wild type. 
As the mutation rate $q$ is raised, the proportion of the wild type $w^*$ present
in the population decreases. When $q$ reaches the error threshold, the wild type
disappears from the population. More precisely the proportion of the wild type becomes comparable to
the proportions of the other genotypes. This is the error catastrophe: 
the selective advantage of the wild type $w^*$ is annihilated by the mutations.

This kind of equilibrium was discovered within the framework
of Eigen's model and was called a quasispecies \cite{EI1,ECS1}, as opposed to 
a species,  which 
refers to an homogeneous solution in chemistry. In fact,
Eigen's model is a model for the evolution of a population of macromolecules governed by
a system of chemical reactions and the laws of kinetics yield a differential system of 
equations whose equilibrium is described by the quasispecies equation.
This system was historically analyzed with approximation and expansion techniques, which in the
asymptotic regime~\eqref{iasym} led to the discovery of the error threshold.

Although
the original goal of Eigen was to understand the first stages of life on Earth,
the notion of quasispecies and the error threshold 
had a profound impact on the understanding of molecular evolution \cite{dope}.
Indeed,
many living organisms seem to satisfy approximately the scaling relation
$$\text{mutation rate}\quad\sim\quad
\frac{1}{\text{genome length}}\,.$$
Unfortunately, for complex creatures, it is very complicated to estimate the mutation rate, 
which is usually extremely small. Viruses, however, have a rather high mutation rate, 
and the orders of magnitude of their genome length and their mutation rate is compatible with
this scaling law.
Moreover, 
some RNA viruses, like the HIV,
evolve with a high mutation rate which 
seems to be close to an error threshold \cite{dope}. 
Why is that so?

In order to survive a virus should achieve two goals.
First, its genetic information should be preserved from one generation to another,
hence its mutation rate has to be below the error threshold.
Second, it has to escape the attacks of the immune system of its hosts, and to do so,
it should mutate as fast as possible. The most efficient strategy is therefore to 
adjust the mutation rate to the left of the error threshold:
this will achieve simultaneously a huge genetic variability,
and the preservation of the pertinent genetic information across generations.
Some promising antiviral strategies, called lethal mutagenesis, consist in using mutagenic drugs which increase 
slightly the mutation rate of the virus in order to
induce an error catastrophe \cite{ADL,CCA, dope}.

Most of the rigorous mathematical analysis of Eigen's model deals with the sharp peak
landscape, or with landscapes presenting a lot of symmetries (see~\cite{bnasl} for
the analysis of a
permutation invariant fitness landscape
and
\cite{bnasr} for a review of the literature).
\subsection{Extension to a general landscape}
We try to extend the formulas obtained in section~\ref{prof} to a general fitness landscape.
Here we work with the genotype space $E=\cA^\ell$
where $\cA$ is a finite alphabet of cardinality $\kappa$. 
We
make the following hypothesis
on the fitness function $f$.
\begin{hyp}
\label{exhyp}
The fitness function $f$ is larger or equal than~$1$, and it is not identically equal to $1$.
\end{hyp}
\noindent
In theorem~\ref{relationFH},
we obtain an interesting equation
involving the covariance of the fitness and the sum of an arbitrary function~$g$ over the letters of the sequences
in the quasispecies distribution
This formula is a considerable generalization of the formula~\eqref{imeanHamm}.
\begin{theorem}
  \label{relationFH}
Let $\cA$ be a finite alphabet of cardinality $\kappa$ and 
	let $E=\cA^\ell$.
Let $g$ be a function defined on $\cA$ with values in $\R$
and let $G:E\to\R$ be defined by
$$\forall u\in E\qquad G(u)\,=\,\sum_{i=1}^\ell g(u(i))\,.$$
For any fitness function $f$ over the genotype set~$E$, 
which is larger or equal than~$1$, and which is not identically equal to $1$, we have
\smallskip
\begin{equation}
    \label{simp}
	\of\,\oG\,=\,\rho\,\ofG+\ell(1-\rho)\of\,\bg\,,
\end{equation}
	where 
	$\rho$ is given in~\eqref{fdefr} and
\begin{gather*}
    \ofG
	\,=\,\sum_{u\in E}f(u)\,G(u)\,x(u)
\,,\quad \bg\,=\,
	\frac{1}{\kappa}
	\sum_{a\in\cA}g(a)
	\,,\cr
\of\,=\,\sum_{u\in E}f(u)x(u)\,,\quad
\oG\,=\,\sum_{u\in E}G(u)x(u)
	\,.
\end{gather*}
\end{theorem}
\noindent
Let us look at the specific case of the sharp peak landscape.
We take $\kappa=2$, $E=\lbrace\,0,1\,\rbrace^\ell$, and
the fitness function defined in formula~\eqref{sp}. For the function $g$, we take 
the identity function on 
$\lbrace\,0,1\,\rbrace$, so that, for any $u\in E$, the value $G(u)$ is
simply the Hamming class of $u$ 
(that is, the number of ones present in $u$). 
In particular, we have
$\ofG\,=\,\oG$ and $\bg=1/2$, so that
the formula~\eqref{simp} becomes
\begin{equation}
    \label{hmp}
	\of\,\oG\,=\,\rho\,\oG+\frac{\ell}{2}(1-\rho)\of\,,
\end{equation}
which is precisely the formula~\eqref{imeanHamm} that
we obtained previously in this context.
We can also consider the neutral case. If the fitness function is constant, then
the quasispecies distribution is in fact the uniform distribution over the genotype space,
so that~$\oG=\ell/2$, and the above relation~\eqref{hmp} still holds. This is not too surprising,
because it amounts to the case where the fitness~$\sigma$ of the 
wild type $w^*$ is equal to $1$.

The proof of theorem~\ref{relationFH} is done
in section~\ref{ext}, where we perform also several
useful computations valid for a general state space~$E$.
The various formulas obtained in section~\ref{ext}
shed new insight into the classical
phenomenon of the error threshold and
the notion of quasispecies, which are discussed in section~\ref{ert}.
Finally, in section~\ref{ext}, we outline a robust strategy
to analyze more general fitness landscapes. This strategy will
be used to study the fitness landscapes introduced in 
the following subsections.
\subsection{Finitely many peaks}
Now that we have well understood the case of the sharp peak landscape, we are ready to consider more complex landscapes.
Here we work with the genotype space $E=\cA^\ell$
where $\cA$ is a finite alphabet of cardinality $\kappa$. 
We
make the following hypothesis
on the fitness function $f$.
\begin{hyp}
	There exist a fixed integer $k$  
	and $k$ fixed values $\sigma_1,\dots,\sigma_k$ such that:
	for any value of $\ell,q$, there exist $k$
	peaks $w_1^*,\cdots,w_k^*$ in $E$
	(whose location may vary with $\ell$ and $q$)
	such that
	the fitness function $f_{\text{FP}}$ is given by
\begin{equation}
\label{fmp}
\forall\,u\in E\qquad
	f_{\text{FP}} (u)\,=\,\begin{cases}
\quad\sigma_i&\quad\text{if}\quad u=w^*_i\,,
\quad 1\leq i\leq k\,,\\
\quad1&\quad\text{otherwise}\,.
\end{cases}
\end{equation}
\end{hyp}
\noindent
We consider again the long chain regime
\begin{equation*}
\ell\to+\infty\,,\qquad
q\to 0\,,\qquad
\ell q\to a\in\,[0,+\infty]\,.
\end{equation*}
To simplify the proofs, we suppose that $a<+\infty$.
Indeed, the case $a=+\infty$
should be considered separately.
We shall prove the following theorem.
%
\begin{theorem}[Error threshold for finitely many peaks]
  \label{ifrrt}
  Let us set
	$$\sigma\,=\,\max_{1\leq i\leq k} \sigma_i\,.$$
	Let $\lambda_{\text{FP}}(\ell,q)$ be the mean fitness for the model
	associated with the fitness function 
	$f_{\text{FP}}$ and the parameters $\ell,q$.
We have the following dichotomy:

\noindent
	$\bullet$ If $a>\ln\sigma$, then
	$\lambda_{\text{FP}}(\ell,q)
\rightarrow 1$.

\noindent
$\bullet$ If $a<\ln \sigma$, then
	$\lambda_{\text{FP}}(\ell,q)
\rightarrow \sigma e^{-a}>1$.
\end{theorem}

\noindent
We see that the situation is very similar to the case of the sharp peak landscape. The critical parameter
depends now on the fitness of the highest peak. 
The conclusion is that, when there is
a finite number of peaks, the lower peaks do not really
interfere with the highest peak.
Two proofs of 
theorem~\ref{ifrrt} 
are presented in section~\ref{fmape}.
\subsection{Plateau}
We have seen that a finite number of small peaks do not interact significantly with the highest peak.
We consider here a fitness landscape
which contains a kind of plateau, that is a large subset of genotypes having a fitness $\sigma>1$.
More precisely, 
we consider the genotype space 
of binary sequences 
$E=\lbrace\,0,1\,\rbrace^\ell$
and the usual mutation kernel $M$.
To simplify the formulas, we suppose that the length $\ell$ is even.
The Hamming class number $H(u)$ of $u\in
\lbrace\,0,1\,\rbrace^\ell$ is the number of ones present in $u$, that is
$$H(u)\,=\,\sum_{1\leq i\leq\ell}u(i)\,.$$
We consider 
the fitness function 
	$f_{\text{PL}}$ defined as follows.
\begin{hyp}
The fitness function 
	$f_{\text{PL}}$
	is given by
\begin{equation}
\label{fplat}
\forall\,u\in 
\lbrace\,0,1\,\rbrace^\ell
	\qquad
	f_{\text{PL}} (u)\,=\,\begin{cases}
	\quad\sigma&\quad\text{if }H(u)=\ell/2\,,\\
\quad1&\quad\text{otherwise}\,.
\end{cases}
\end{equation}
\end{hyp}
\noindent
The plateau associated with the fitness function
$f_{\text{PL}} (u)$ consists of the sequences
having $\ell/2$ zeroes and $\ell/2$ ones.
In particular, it contains sequences which are very far
away for the Hamming distance.
This might go against the intuitive notion of a plateau, 
especially in low-dimensional spaces, where we would 
usually take for a plateau a closed neighborhood 
around a fixed connected set.
However, our goal is to find a subset of 
$\lbrace\,0,1\,\rbrace^\ell$ which is very stable with 
respect to the mutations, and 
the plateau of
$f_{\text{PL}} (u)$ is the simplest example of such a set.
We could also modify this set so that it is closer to our
intuition of a plateau, by adding to it all the sequences
which are at Hamming distance less than $\sqrt{\ell}$ from it.
All the results presented here and in the subsequent section
would hold for this larger set as well (yet the proofs would
become more complicated).

We consider again the long chain regime with $a$ finite:
\begin{equation*}
\ell\to+\infty\,,\qquad
q\to 0\,,\qquad
\ell q\to a\in\,[0,+\infty[\,.
\end{equation*}
\begin{theorem}
  \label{plart}
	Let $\lambda_{\text{PL}}(\ell,q)$ be the mean fitness for the model
	associated with the fitness function 
	$f_{\text{PL}}$ and the parameters $\ell,q$.
	For any $a>0$, we have 
	$$\lim_{ \genfrac{}{}{0pt}{1}{\ell\to\infty,\, q\to 0 } {{\ell q} \to a } }
	\lambda_{\text{PL}}(\ell,q)
	\,=\,
	\lambda(a,\sigma)\,,$$
	where $\lambda(a,\sigma)$ is the unique real number $\lambda>1$ such that
\begin{equation}
	\label{fiteq}
	\frac{1}{\sigma-1}\,=\,
	\sum_{n\geq 1}
	\frac{e^{-an}}{\lambda^n}
	\sum_{k\geq 0}
	\frac{(an)^{2k}}{2^{2k}(k!)^2}\,.
\end{equation}
%
\end{theorem}
\noindent
The situation for the fitness landscape
	$f_{\text{PL}}$
	with the huge plateau is very different from the sharp peak landscape.
	Indeed, the error threshold phenomenon described in
theorem~\ref{ierrt} does not occur, and there is no phase transition in the long chain regime.
It turns out that, if we shift the plateau away from $\ell/2$, then
an error threshold phenomenon reappears.
We consider 
the fitness function 
	$f_{\text{PL}}$ defined as follows.
\begin{hyp}
	Let $\sigma>1$ and $\alpha\in [0,1]$.
The fitness function 
	$f_{\text{PL}}$
	is given by
\begin{equation}
\label{faplat}
\forall\,u\in 
\lbrace\,0,1\,\rbrace^\ell
	\qquad
	f_{\text{PL}} (u)\,=\,\begin{cases}
	\quad\sigma&\quad\text{if }H(u)=\lfloor \alpha\ell\rfloor\,,\\
\quad1&\quad\text{otherwise}\,,
\end{cases}
\end{equation}
where
	$\lfloor \alpha\ell\rfloor$ is the integer part of $\alpha\ell$.
\end{hyp}
\noindent
The cases $\alpha=0$ and $\alpha=1$ correspond in fact to the
sharp peak landscape, which has already been studied. 
We shall therefore exclude these cases.
\begin{theorem}
  \label{palart}
  Suppose that $\alpha\not\in\{\,0,1/2,1\,\}$. 
	Let $\lambda_{\text{PL}}(\ell,q)$ be the mean fitness for the model
	associated with the fitness function 
	$f_{\text{PL}}$ and the parameters $\ell,q$.
	There exist two positive values $a^1_c(\sigma,\alpha)\leq a^2_c(\sigma,\alpha)$ such that:
	if $a\geq a^2_c(\sigma,\alpha)$, then
	$$\lim_{ \genfrac{}{}{0pt}{1}{\ell\to\infty,\, q\to 0 } {{\ell q} \to a } }
	\lambda_{\text{PL}}(\ell,q)
	\,=\,1\,.$$
	If $a<a^1_c(\sigma,\alpha)$, then
	$$\liminf_{ \genfrac{}{}{0pt}{1}{\ell\to\infty,\, q\to 0 } {{\ell q} \to a } }
	\lambda_{\text{PL}}(\ell,q)
	\,>\,1\,.$$
\end{theorem}
\noindent
Of course, it would be more satisfactory to know that $a^1_c(\sigma,\alpha)=a^2_c(\sigma,\alpha)$.
This would be a consequence of the 
monotonicity of the function~$\phi_\alpha(a)$ defined in~\eqref{phial}, 
unfortunately
this fact does not seem so obvious.

The proofs of 
theorems~\ref{plart} 
and~\ref{palart} 
are presented in section~\ref{late}.
The reason underlying these two results is the following. In the neutral fitness landscape,
the most likely Hamming class is already the class $\ell/2$. By installing a plateau precisely
on this Hamming class, we reinforce its stability. When we shift the plateau away from the 
Hamming class $\ell/2$, we create a competition between two Hamming classes, and this leads to
the occurrence of a phase transition.
\subsection{Survival of the flattest}
We finally prove rigorously within our framework the possible occurrence of
the survival of the flattest, a phenomenon which has been previously discovered by
	Wilke, Wang, Ofria, Lenski and Adami \cite{Wixa}
	and which has been investigated in several works 
	\cite{CFJS,Frankl,FLAT,teja}.
	We state next the precise mathematical result we obtain.
We will now combine the sharp peak landscape with a plateau.
We consider the genotype space 
of binary sequences 
$E=\lbrace\,0,1\,\rbrace^\ell$
and the usual mutation kernel $M$, with 
the following fitness function.
To simplify the formulas, we suppose that the length $\ell$ is even.
\begin{hyp}
Let $\delta,\sigma\geq 1$ and let $f_{\text{SP/PL}}$
	be the
	fitness function given by
\begin{equation}
\label{fsurv}
\forall\,u\in 
\lbrace\,0,1\,\rbrace^\ell
	\qquad
	f_{\text{SP/PL}}(u)\,=\,\begin{cases}
	\quad\delta&\quad\text{if }u=0\cdots 0\,,\\
	\quad\sigma&\quad\text{if }u\text{ contains }
	\frac{\ell}{2}\text{ zeroes and }
	\frac{\ell}{2}\text{ ones}
	\,,\\
\quad1&\quad\text{otherwise}\,.
\end{cases}
\end{equation}
\end{hyp}
\begin{theorem}
  \label{isurt}
  Let $a>0$ and let
	$\lambda(a,\sigma)$ be the unique real number $\lambda>1$ such that
\begin{equation}
	\label{rfiteq}
	\frac{1}{\sigma-1}\,=\,
	\sum_{n\geq 1}
	\frac{e^{-an}}{\lambda^n}
	\sum_{k\geq 0}
	\frac{(an)^{2k}}{2^{2k}(k!)^2}\,.
\end{equation}
	Let $\lambda_{\text{SP/PL}}(\delta,\sigma,\ell,q)$ be the mean fitness for the model
	associated with the fitness function 
	$f_{\text{SP/PL}}$ and the parameters $\ell,q$.
	We have the following
convergence in the long chain regime:
	$$\lim_{ \genfrac{}{}{0pt}{1}{\ell\to\infty,\, q\to 0 } {{\ell q} \to a } }
	\lambda_{\text{SP/PL}}(\delta,\sigma,\ell,q)
	\,=\,\max\big(\delta\exa,\lambda(a,\sigma)\big)\,.$$
	Let us denote by $y(0)$ the fraction of the sequence $0\cdots 0$ 
	in the population
	at equilibrium, and by
	$y(\ell /2)$ the fraction of the sequences
	having $\ell/2$ zeroes and 
	and $\ell/2$ ones
	in the population
	at equilibrium.

\noindent
$\bullet$ If $\lambda(a,\sigma)>\delta\exa$, then $y(0)\to 0$,
	$y(\ell/2)\to \displaystyle\frac{\lambda(a,\sigma)-1}{\sigma-1}$.

\noindent
$\bullet$ If $\delta\exa>\lambda(a,\sigma)$, then $y(\ell/2)\to 0$,
	$y(0)\to\displaystyle \frac{\delta\exa-1}{\delta-1}$.
\end{theorem}
\noindent
A plateau is always more stable than a peak of the same height,
that is, for any $a>0$, any $\sigma>1$, we have
$\lambda(a,\sigma)>\sigma\exa$.
However,
for a fixed value $a>0$, there exist positive values
$\sigma<\delta$ such that
$\lambda(a,\sigma)>\delta\exa$. Indeed,
let us compute the right-hand member of equation~\eqref{rfiteq}
where $\lambda$ is replaced by $\delta\exa$:
\begin{align*}
	\sum_{n\geq 1}
	\frac{e^{-an}}{\big(\delta\exa\big)^n}
	\sum_{k\geq 0}
	\frac{(an)^{2k}}{2^{2k}(k!)^2}&\,=\,
	\sum_{n\geq 1}
	\frac{1}{\delta^n}
	\sum_{k\geq 0}
	\frac{(an)^{2k}}{2^{2k}(k!)^2}\cr
				      &\,\geq \,
	\sum_{n\geq 1}
	\frac{1}{\delta^n}
	\Big(1+ \frac{(an)^{2}}{4}\Big)
	\,=\,
	\frac{1}{\delta-1}+
	\frac{a^{2}}{4}\frac{\delta(\delta+1)}{(\delta-1)^3}
	\,.
\end{align*}
So, for any value of $\delta>\sigma$ such that
	$$
	\frac{1}{\sigma-1}\,<\,\frac{1}{\delta-1}+
	\frac{a^{2}}{4}\frac{\delta(\delta+1)}{(\delta-1)^3}\,,$$
	we will indeed have 
$\lambda(a,\sigma)>\delta\exa$.
For such values,
the quasispecies will be located on the plateau, despite
the fact that the height $\sigma$ of the plateau is strictly
less than the height of the peak.
In this situation, we witness the survival of the flattest.

Theorem~\ref{isurt} is proved in
section~\ref{sof}.
The route to prove theorems~\ref{ifrrt} and~\ref{isurt} is quite long. It rests on the exact 
formulas proved in subsection~\ref{commn} and 
the general strategy developed in section~\ref{ext}.

\section{The sharp peak landscape}
\label{prof}

Our first goal is to develop an exact formula for the quasispecies
on the sharp peak landscape. All our subsequent study of the sharp peak
landscape will rely on this formula.
\subsection{A representation formula for the quasispecies}
The computations performed in this subsection are in fact valid without
any assumption on the geometry of the genotype space~$E$.
We merely suppose that $E$ is finite and that the coefficients of
the matrix $M$ are all positive.
Suppose that $(x(u), u\in E)$ is a solution of the system (\ref{qs})
satisfying the constraint (\ref{cons}).
Let $\lambda$ be the mean fitness of this solution, defined by 
\begin{equation}
\label{meanfit}		
\lambda\,=\,\sum_{v\in E}x(v)f(v)\,=\,\sigma x(w^*)+1-x(w^*)\,.
\end{equation}
Obviously, the mean fitness $\lambda$ satisfies $1\leq\lambda\leq\sigma$
and $\lambda=1$ if and only if $x(w^*)=0$. 
By hypothesis, all the coefficients
of the mutation matrix $M$ are positive, and it follows from equation (\ref{qs}) that
\begin{equation}
\label{posit}		
\forall\, u\in E\qquad
x(u)\,\sigma\,\geq\,
\,\sum_{v\in E}x(v)f(v)M(v,u)\,>\,0\,.
\end{equation}
Therefore $x(u)>0$ for all $u\in E$.
The equation (\ref{qs}) for a generic genotype $u$ reads:
\begin{equation}
\label{generic}
\lambda
x(u)\,=\,
{\sigma} x(w^*)M(w^*,u)+
\sum_{v\neq w^*}x(v)M(v,u)\,.
\end{equation}
We rewrite this equation as
\begin{equation}
\label{genericb}
x(u)\,=\,
\frac{\sigma-1}{\lambda} x(w^*)M(w^*,u)+
\frac{1}{\lambda} 
\sum_{v\in E}x(v)M(v,u)\,.
\end{equation}
Using this very formula for $x(v)$, 
we replace $x(v)$ in the sum, thereby getting
\begin{multline*}
x(u)\,=\,
\frac{\sigma-1}{\lambda} x(w^*)M(w^*,u)+
\frac{\sigma-1}{\lambda^2} 
\sum_{v\in E} x(w^*) M(w^*,v) M(v,u)\cr
	+ \frac{1}{\lambda^2} \sum_{v,v'\in E}x(v')M(v',v)M(v,u)\,.
\end{multline*}
We iterate this process. After $N-1$ steps, we get
\begin{multline*}
x(u)\,=\,
\sum_{1\leq n\leq N}
\frac{\sigma-1}{\lambda^n} 
\sum_{v_1,\dots,v_{n-1}\in E}
	\kern-7pt
	x(w^*)
M(w^*,v_1)\cdots M(v_{n-1},u)\cr
\,+\,
\frac{1}{\lambda^N} 
\sum_{v_1,\dots,v_{N}\in E} \kern-7pt
	x(v_1)
M(v_1,v_2)\cdots M(v_{N},u)\,.
\end{multline*}
The sums of products involving the matrix $M$ have a natural
probabilistic interpretation.
Let $(X_n)_{n\in\mathbb N}$ be the Markov chain on~$E$ with 
transition matrix~$M$. 
The previous formula for $x(u)$ can now be rewritten as
\begin{multline*}
x(u)\,=\,
\sum_{1\leq n\leq N}
\frac{\sigma-1}{\lambda^n} 
	x(w^*)
\PP\big(X_{n}=u\,\big|\,X_0=w^*\big)\cr
\,+\,
\frac{1}{\lambda^N} 
\sum_{v_1\in E}
	x(v_1)
\PP\big(X_{N}=u\,\big|\,X_0=v_1\big)\,.
\end{multline*}
By (\ref{meanfit}) and~\eqref{posit}, the mean fitness $\lambda$ is strictly larger than one,
therefore
the last term goes to $0$ as $N$ goes to $\infty$. 
Sending $N$ to $\infty$, and using the fact that
\begin{equation}
\label{lsi}
\lambda-1\,=\,(\sigma-1)x(w^*)\,,\end{equation}
we get the following result.
\begin{proposition}
Any solution $(x(u),u\in E)$ of the quasispecies equations (\ref{qs})
for the sharp peak function $f_{SP}$ defined in~\eqref{sp}
satisfies
\begin{equation}
\label{probsol}
x(u)\,=\,
\sum_{n\geq 1}
\frac{\lambda-1}{\lambda^n} 
\PP\big(X_{n}=u\,\big|\,X_0=w^*\big)\,,
\end{equation}
where $(X_n)_{n\geq0}$ is the Markov chain on $E$
with transition matrix $M$.
\end{proposition}

Let us take $u=w^*$ in  formula (\ref{probsol}). 
Using again~\eqref{lsi}, we obtain
\begin{equation}
\label{lambda}
	\frac{1}{\sigma-1}
\,=\, 
\sum_{n\geq 1}
\frac{1}{\lambda^n} 
\PP\big(X_{n}=w^*\,\big|\,X_0=w^*\big)\,.
\end{equation}
The right-hand side of this equation is a continuous decreasing function of $\lambda$,
which is equal to $+\infty$ when $\lambda=1$ and is less than or equal to $1/(\sigma-1)$
when $\lambda=\sigma$. Therefore there exists exactly one value of $\lambda$ in 
$[1,\sigma]$ which satisfies this equation. 
Our next goal is to obtain an explicit 
expression for
$P\big(X_{n}=w^*\,\big|\,X_0=w^*\big)$ and to perform the summation of the series 
appearing in~(\ref{lambda}). To do so, we shall make additional assumptions on
the geometry of the fitness landscape.
\subsection{Mutation dynamics}
\label{mutdyn}
We now turn our attention to
the genotype space $E=\cA^\ell$
where $\cA$ is a finite alphabet of cardinality $\kappa$ and we consider
the mutation scheme~\eqref{mut}. 
In this model, the mutations occur independently at each site. An important consequence
of this structural assumption is that the components 
$(X_n(i),1\leq i\leq\ell)$
of the Markov chain $X_n$
are themselves independent Markov chains with state space $\cA$
and transition matrix
$$\begin{pmatrix}
\displaystyle 1-q & \displaystyle \frac{q}{\kappa-1} &\cdots & \displaystyle \frac{q}{\kappa-1}\cr
\displaystyle \frac{q}{\kappa-1}& 1-q &\ddots & \displaystyle \frac{q}{\kappa-1}\cr
\vdots &\ddots & \ddots &\vdots\cr
\displaystyle \frac{q}{\kappa-1}& \cdots & \displaystyle \frac{q}{\kappa-1} & 1-q\cr
\end{pmatrix}\,.
$$
The non-diagonal terms in this matrix are all equal to $q/(\kappa-1)$.
Since we want to compute
$\PP\big(X_{n}=w^*\,\big|\,X_0=w^*\big)$, 
we shall lump together the letters which
differ from the wild type $w^*$, and we shall do this for each component.
More precisely, 
we define a process
$(V_n)_{n\geq 0}$ by setting
$$\forall i\in\ul\qquad V_n(i)\,=\,
\begin{cases}
0 &\text{if $X_n(i)=w^*(i)$}\,,\\
1 &\text{if $X_n(i)\neq w^*(i)$}\,.\\
\end{cases}
$$
The binary word $V_n$ indicates the sites where $X_n$ and $w^*$ differ.
In particular, $V_n$ is a deterministic function of $X_n$.
Thanks to the specific form of the transition matrix, it turns out that
$V_n(i)$ is still a Markov chain.
This is a particular case of 
the lumping theorem of Kemeny and Snell 
(see theorem~$6.3.2$ of \cite{KS}). To see why it is so,
let us consider for instance
the first two steps of the first component $V_n(1)$ and let us compute
\begin{multline*}
\PP(V_2(1)=0,\,
V_1(1)=1
\,\big|\,V_0(1)=0\big)\cr
	\,=\,
\sum_{x_1\in \cA\setminus \{\,w^*(1)\,\} }
	\PP(X_2(1)=w^*(1),\,
	X_1(1)=x_1
\,\big|\,V_0(1)=0\big)
	\cr
	\,=\,
\sum_{x_1\in \cA\setminus \{\,w^*(1)\,\} }
	\PP(X_2(1)=w^*(1)\,|\,
	X_1(1)=x_1)
	\cr
	\times
\PP( X_1(1)=x_1 \,|\,
X_0(1)=w^*(1)\big)
	\,=\, (\kappa-1)\times\frac{q}{\kappa-1}\times \frac{q}{\kappa-1}
	\,=\, {q}\frac{q}{\kappa-1}\,.
\end{multline*}
The simplification in the final result comes from the fact that the probabilities to find or to lose
a letter from the wild type $w^*$ are the same for all the letters in
$\cA\setminus \{\,w^*(1)\,\}$. A similar computation can be done for a finite number of 
steps. We conclude that
$V_n(i)$ is 
the two states Markov chain
that we define and study next.
Let $(E_n)_{n\geq 0}$
be the Markov chain 
with state space $\{0,1\}$ and transition
matrix
$$T\,=\,\begin{pmatrix}
\displaystyle 1-q & q \cr
	\displaystyle \frac{q}{\kappa-1}& 1-\displaystyle \frac{q}{\kappa-1}\cr
\end{pmatrix}\,.
$$
The eigenvalues of $T$ are $1$ and 
\begin{equation}
	\label{ro}
\rho\,=\,
1-\displaystyle \frac{\kappa q}{\kappa-1}\,.
\end{equation}
We will study some asymptotic regimes where $q$ is sent to~$0$,
so we can suppose that $q\leq 1-1/\kappa$ 
(otherwise $\rho$ might be negative and this creates additional complications).
It is a standard exercise to compute the powers of $T$:
$$\forall n\geq 1\qquad T^n\,=\,
\begin{pmatrix}
\displaystyle \frac{1}{\kappa} + 
\displaystyle \frac{\kappa-1}{\kappa}\rho^n & 
\displaystyle \frac{\kappa-1}{\kappa} \Big(1-\rho^n\Big)  \cr
\cr
\displaystyle 
\displaystyle \frac{1}{\kappa}\Big(1-\rho^n \Big)& 
\displaystyle \frac{\kappa-1}{\kappa}
+\displaystyle \frac{1}{\kappa}
\rho^n  \cr
\end{pmatrix}\,.
$$
Here is a simple illuminating way to realize the
Markov chain 
$(E_n)_{n\geq 0}$ and to understand the expression of the $n$-th power $T^n$.
Let
$(\varepsilon_n)_{n\geq 1}$ be an i.i.d. sequence of Bernoulli random variables with
parameter $\rho$.
Suppose that $E_{n-1}=e\in \zu$.
If $\varepsilon_n=1$, then we set $E_n=e$.
If $\varepsilon_n=0$, then we choose a letter uniformly over $\cA$,
independently of the past history until time~$n$, and we set $E_n=0$ if the chosen letter
is the one of the wild type and $E_n=1$ otherwise. Now, the event $E_n=E_0$ can occur
in two different ways.
Either $\varepsilon_1=\cdots=\varepsilon_n=1$, or one of the
$\varepsilon_1,\dots,\varepsilon_n$ is zero, in which case $E_n=0$ with probability $1/\kappa$
and 
$E_n=1$ with probability $(\kappa-1)/\kappa$, thus 
\begin{align*}
\PP(E_n=0\,|\,E_0=0)\,&=\,\rho^n+\Big(1-\rho^n\Big)\frac{1}{\kappa}\,,\cr
\PP(E_n=1\,|\,E_0=1)\,&=\,\rho^n+\Big(1-\rho^n\Big)\frac{\kappa-1}{\kappa}\,,
\end{align*}
and we recover the expression of the diagonal coefficients of $T^n$.
In words, the status $E_n$ at step $n$ is the same as at time $0$ if no mutation
has occurred, or if the last mutation results in the same letter as the wild type (case $E_0=0$) or in
a different letter (case $E_0=1$).
Similarly, the event
$E_n\neq E_0$ can occur only if one of the
$\varepsilon_1,\dots,\varepsilon_n$ is zero, and the last mutation event yields the
adequate letter,
thus
\begin{align*}
\PP(E_n=1\,|\,E_0=0)\,&=\,\Big(1-\rho^n\Big)\frac{\kappa-1}{\kappa}\,,\cr
\PP(E_n=0\,|\,E_0=1)\,&=\,\Big(1-\rho^n\Big)\frac{1}{\kappa}\,.
\end{align*}
Now the probability
$\PP\big(X_{n}=w^*\,\big|\,X_0=w^*\big)$ can be rewritten with the help of $V_n$ and $E_n$ as
\begin{align}\label{comput}
	\PP\big(X_{n}=w^*&\,\big|\,X_0=w^*\big)\,=\,
\PP\big(X_{n}(i)=w^*(i), \,1\leq i\leq\ell\,\big|\,X_0=w^*\big)\cr
	 &\,=\,
\PP\big(V_{n}(i)=0, \,1\leq i\leq\ell\,\big|\,X_0=w^*\big)\cr
	 &\,=\,
	{\prod_{1\leq i\leq\ell}} \PP\big(V_{n}(i)=0\,\big|\,X_0=w^*\big)\cr
	 &\,=\,
\Big(\PP\big(E_{n}=0\,\big|\,E_0=0\big)\Big)^\ell
	\,=\,
	\Big( \displaystyle \frac{1}{\kappa} + 
	\displaystyle \frac{\kappa-1}{\kappa}\rho^n  \Big)^\ell\,.
\end{align}
\subsection{The equation for the mean fitness}
We have all the material necessary to obtain the 
equation~\eqref{innlambda}.
Substituting the expression obtained in (\ref{comput}) into formula~(\ref{lambda}),
we obtain
\begin{equation}
\label{nlambda}
	\frac{1}{\sigma-1}
\,=\, 
\sum_{n\geq 1}
\frac{1}{\lambda^n} 
	\Big( \displaystyle \frac{1}{\kappa} + 
	\displaystyle \frac{\kappa-1}{\kappa}\rho^n  \Big)^\ell\,.
\end{equation}
We expand the power in $\ell$, we exchange the summations and we re-sum the series as follows:
\begin{align*}
	\frac{1}{\sigma-1}
&\,=\, 
\sum_{n\geq 1}
\frac{1}{\lambda^n} 
	\sum_{0\leq k\leq \ell}
	\binom{\ell}{k}
	\Big( \displaystyle \frac{1}{\kappa}\Big)^{\ell-k}  
	\Big(\displaystyle \frac{\kappa-1}{\kappa}\rho^n  \Big)^{k}\cr
&\,=\, \displaystyle \frac{1}{\kappa^\ell}
	\sum_{0\leq k\leq \ell}
	\binom{\ell}{k}
	(\kappa-1)^k
\sum_{n\geq 1}
	 \frac{ \rho^{kn}}{\lambda^n}\cr
& \,=\, \displaystyle \frac{1}{\kappa^\ell}
	\sum_{0\leq k\leq \ell}
	\binom{\ell}{k}
	(\kappa-1)^k
	\frac{1}{
	\displaystyle
		\frac{\lambda}{\rho^{k}}-1}
	\,.
\end{align*}
Thus the equation satisfied by the mean fitness $\lambda$ reads:
\begin{equation}
\label{nalambda}
	\frac{1}{\sigma-1}
\,=\, \displaystyle \frac{1}{\kappa^\ell}
	\sum_{0\leq k\leq \ell}
	\binom{\ell}{k}
	(\kappa-1)^k
	\frac{1}{
	\displaystyle
		\frac{\lambda}{\rho^{k}}-1}
	\,.
\end{equation}
This equation was discovered by
Bratus, Novozhilov and Semenov,
see formulas~$(5.5)$ and $(5.6)$ in \cite{BSN1},
they exploited this equation to derive rigorous bounds on the mean fitness.
This formula calls naturally for a probabilistic interpretation. 
Let $S_\ell$ be a random variable with distribution the binomial law with parameters $\ell$ and $(\kappa-1)/\kappa$.
Denoting by $\EE$ the expectation,
equation~(\ref{nalambda}) can be rewritten as
\begin{equation}
\label{nnlambda}
	\frac{1}{\sigma-1}
	\,=\, \EE\left(
	\frac{1}{
	\displaystyle
		\frac{\lambda}{\rho^{S_\ell}}-1}
	\right)\,.
\end{equation}
Our analysis of the quasispecies equations on the sharp peak landscape rests on the identity~\eqref{nlambda} or its equivalent form~\eqref{nnlambda}.
Indeed, if we manage to estimate the mean fitness $\lambda$, then we will also have an estimate on the proportion $x(w^*)$
of the wild type present in the population. Moreover, once we know $\lambda$ or $x(w^*)$,
the proportions of the other types are completely determined by formula~\eqref{probsol}.
\subsection{An exact formula for the quasispecies}
\label{exqs}
Loosely speaking, a quasispecies is
a cloud of mutants centered around a specific genotype.
The structure of this cloud depends on the parameters $\sigma$ and $q$. 
In fact,
the proportion of the wild type $w^*$
decreases as the mutation rate $q$ increases, and it becomes comparable to the proportions
of the other genotypes when $q$ reaches
the error threshold $q^*$. This fascinating phenomenon will be proved rigorously
in section~\ref{ert}. When $q$ is a little below $q^*$, we observe
a cloud of mutants centered around $w^*$ which contains a very small proportion of
the wild type $w^*$, the vast majority of the genotypes present in the population 
will differ from $w^*$. Yet the genetic information carried out by $w^*$ is still
present in the population and it determines its structure. 
This paradoxical situation has led several biologists
to argue that the selection operates at the level of the quasispecies, and not
at the level of individuals \cite{dope}. 
Thus an important goal is
to understand better the statistical composition of the cloud of mutants.
In order to do so, we classify the genotypes according to their Hamming distance
to the wild type $w^*$. 
For $k\geq 1$, we define 
the Hamming class $k$ as the set $\cH_k$ of the genotypes which differ from $w^*$
at exactly $k$ indices, i.e.,
\begin{equation}
	\cH_k\,=\,\big\{\,u\in E :d(u,w^*)=k\,\big\}\,.
\end{equation}
We shall exploit
further formula~\eqref{probsol} in order to derive an
exact formula
for the proportion 
${\cQ(k)}$
of the genotypes belonging to the Hamming class $k$. For $k\geq 1$, we have,
thanks to formula~\eqref{probsol},
\begin{align}
\label{crobsol}
	{\cQ}(k)
	&\,=\,\sum_{u\in\cH_k}
\sum_{n\geq 1}
\frac{\lambda-1}{\lambda^n} 
\PP\big(X_{n}=u\,\big|\,X_0=w^*\big)\,
\cr
	&\,=\,
\sum_{n\geq 1}
\frac{\lambda-1}{\lambda^n} 
	\sum_{u\in\cH_k}
\PP\big(X_{n}=u\,\big|\,X_0=w^*\big)\,
\cr
	&\,=\,
\sum_{n\geq 1}
\frac{\lambda-1}{\lambda^n} 
	\PP\big(X_{n} \in\cH_k
	\,\big|\,X_0=w^*\big)\,.
\end{align}
We have already noticed 
that the components of $X_n$,
$(X_n(i),1\leq i\leq\ell)$, are independent Markov chains.
Therefore, using the notation of section~\ref{mutdyn},
\begin{align}
\label{probaclassek}
	\PP\big(X_{n} \in\cH_k
	&\,\big|\,X_0=w^*\big)\,=\,
	\PP\Big(\sum_{1\leq i\leq\ell}V_{n}(i)=k\,\big|\,X_0=w^*\Big)\cr
	&\,=\,
	\binom{\ell}{k}
	\Big(\PP\big(E_{n}=1\,\big|\,E_0=0\big)\Big)^k
	\Big(\PP\big(E_{n}=0\,\big|\,E_0=0\big)\Big)^{\ell-k}
	\cr
	&\,=\,
	\binom{\ell}{k}
	\Bigg( \Big(1-\rho^n\Big)\frac{\kappa-1}{\kappa} \Bigg)^k
	\Bigg(
\rho^n+\Big(1-\rho^n\Big)\frac{1}{\kappa}
	\Bigg)^{\ell-k}\,.
\end{align}
Recall that
$\rho=
1-{\kappa q}/({\kappa-1})$.
Plugging~\eqref{probaclassek} in~\eqref{crobsol}, we get an exact formula for the quasispecies distribution,
in terms of a series involving $\lambda$ and $\rho$:
\begin{equation}
\label{exactquas}
	{\cQ}(k)
	\,=\,
\sum_{n\geq 1}
\frac{\lambda-1}{\lambda^n} 
	\binom{\ell}{k}
	\Bigg( \Big(1-\rho^n\Big)\frac{\kappa-1}{\kappa} \Bigg)^k
	\Bigg(
\rho^n+\Big(1-\rho^n\Big)\frac{1}{\kappa}
	\Bigg)^{\ell-k}\,.
\end{equation}
We can even compute the sum of the series, by developing the powers, and we get a finite algebraic formula:
\begin{equation}
\label{exactfquas}
	{\cQ}(k)
	\,=\,
(\lambda-1)
	\binom{\ell}{k}
	\frac{(\kappa-1)^k}{\kappa^\ell} 
	\sum_{i=0}^k
	\sum_{j=0}^{\ell-k}
	\binom{k}{i}
	\binom{\ell-k}{j}
	\frac{(-1)^i(\kappa-1)^j\rho^{i+j}}
	{\lambda-\rho^{i+j}}
	\,.
\end{equation}
Of course this formula is complicated, yet it expresses completely the dependence of 
	${\cQ}(k)$ as a function of $\lambda$ and $q$ and it shows the complexity of
	the sharp peak landscape.
Surprisingly, the
mean and the variance of the Hamming distance of the quasispecies distribution have very simple expressions.
Indeed, the mean of the Hamming distance of the distribution 
$\overline{\cQ}$ is given by
\begin{equation}
\label{meanHamm}
	\overline{\cQ} = 
\sum_{k=0}^\ell k \,{\cQ}(k) = 
	\frac{\ell q\, \lambda}{\lambda-\rho}\,.
\end{equation}
To compute this mean, we rely on formula~\eqref{exactquas} to write
\begin{align}
	\overline{\cQ}&\, = \,
\sum_{k=0}^\ell k \,{\cQ}(k)\cr
		      &\,= \,
\sum_{n\geq 1}
\frac{\lambda-1}{\lambda^n} 
\sum_{k=0}^\ell k 
	\binom{\ell}{k}
	\Bigg( \Big(1-\rho^n\Big)\frac{\kappa-1}{\kappa} \Bigg)^k
	\Bigg(
\rho^n+\Big(1-\rho^n\Big)\frac{1}{\kappa}
	\Bigg)^{\ell-k}\,.
\end{align}
In the inner sum, we recognize the expectation of a binomial law, whence
$$\overline{\cQ}  \,=\,
 \sum_{n\geq 1}\frac{ \lambda-1 }{\lambda^n}
 \ell  \Big(1-\rho^n\Big)\frac{\kappa-1}{\kappa} 
 \,.
$$
By summing the geometric series and using the definition of $\rho$, we obtain formula \eqref{meanHamm}.
In fact, all the successive moments of the quasispecies distribution can be computed in this way.
\section{The error threshold}
\label{tet}
We still consider the case of the sharp peak landscape described in section~\ref{dspl}. In the previous section~\ref{prof}, we did
exact computations with the length $\ell$ fixed. Here we shall
perform asymptotic expansions in the long chain regime. We will
revisit the classical error threshold phenomenon and compute the
limiting quasispecies distribution.
\subsection{An upper bound on $\lambda$}
The goal of this subsection is to prepare the ground for the technical
proof of the error threshold result.
We start from the identity~\eqref{nnlambda} and we wish to obtain
an upper bound on $\lambda$. 
From the weak law of large numbers, we know that the most likely values for $S_\ell$ are those
around its mean $(\kappa-1)\ell/\kappa$, so we pick a positive number $\varepsilon$ and
we split the expectation as follows:
\begin{multline}
\label{iambda}
	\frac{1}{\sigma-1}
	\,=\, 
	\sum_{0\leq k\leq \ell}
	\PP(S_\ell=k)
	\frac{1}{
	\displaystyle
		\frac{\lambda}{\rho^{k}}-1}\cr
	\,=\, 
	\sum_{k:|k-(\kappa-1)\ell/\kappa|>\varepsilon\ell}
	\PP(S_\ell=k)
	\frac{1}{
	\displaystyle
		\frac{\lambda}{\rho^{k}}-1}
	+
	\sum_{k:|k-(\kappa-1)\ell/\kappa|\leq\varepsilon\ell}
	\PP(S_\ell=k)
	\frac{1}{
	\displaystyle
		\frac{\lambda}{\rho^{k}}-1}
\,.
\end{multline}
Taking advantage from the fact that $\rho\leq 1$, we bound $\rho^k$ by $1$ in the first sum and by
$\rho^{(\kappa-1)\ell/\kappa-\varepsilon\ell}$ in the second, this leads to the inequality
\begin{equation}
\label{jambda}
	\frac{1}{\sigma-1}
	\,\leq\, 
	\frac{1}{ \displaystyle {\lambda}-1}
	\PP\Big(\Big|S_\ell-
	\frac{\kappa-1}{\kappa}\ell\Big|>\varepsilon\ell\Big)
	+
	\frac{1}{ \displaystyle \frac{\lambda}{
		\rho^{(\kappa-1)\ell/\kappa-\varepsilon\ell} }-1}
\,.
\end{equation}
We 
apply the Chebyshev inequality to the binomial random variable $S_\ell$ and we get the classical estimate:
\begin{equation}
\label{cheb}
	\PP\Big(\Big|S_\ell-
	\frac{\kappa-1}{\kappa}\ell\Big|>\varepsilon\ell\Big)
	\,\leq\, \frac{\text{variance}(S_\ell)}{\varepsilon^2\ell^2}
	\,=\, \frac{\kappa-1}{\kappa^2\varepsilon^2\ell}\,.
\end{equation}
Plugging the Chebyshev inequality into \eqref{jambda}, we conclude that
\begin{equation}
\label{kambda}
\forall\varepsilon>0\qquad
	\frac{1}{\sigma-1}
	\,\leq\, 
	\frac{\kappa-1}{\kappa^2\varepsilon^2\ell(\lambda-1)}
	+
	\frac{1}{ \displaystyle \frac{\lambda}{
		\rho^{(\kappa-1)\ell/\kappa-\varepsilon\ell} }-1}
\,.
\end{equation}
Let us pause for one moment to look at this inequality. Obviously, as $\lambda$ goes to $\infty$, the right-hand side
goes to $0$, and the inequality cannot hold. In other words, for this inequality to hold, $\lambda$ must not be too large.
Thus, from this inequality, we should be able to derive an upper bound on $\lambda$.
To get the best possible upper bound, we could rewrite it as a polynomial inequality of degree two in $\lambda$ and compute
the associated roots, but this leads to messy expressions. Instead, after a few trials (hidden to the reader), we obtain
the following inequality.
\begin{lemma}
The mean fitness $\lambda$ satisfies
\begin{equation}
\label{major}
\lambda
\,\leq\,
	\max\Bigg(1+\frac{4\sigma}{\ell^{1/4}},
\sigma \Big(1-\displaystyle \frac{\kappa q}{\kappa-1}\Big)^{
	{(\kappa-1)\ell}/{\kappa}-\ell^{2/3}}\Big(1+\frac{1}{\ell^{1/13}}\Big)\Bigg)\,.
\end{equation}
\end{lemma}
\begin{proof} Suppose that $\lambda$ is larger than the upper bound in~\eqref{major}. 
	We check that this is not compatible with inequality~\eqref{kambda} with the choice $\varepsilon={\ell}^{-1/3}$. 
	Indeed, we have on one hand
	$$\lambda>1+\frac{4\sigma}{\ell^{1/4}}\qquad\Longrightarrow\qquad
	\frac{1}{\ell^{1/3}(\lambda-1)}\,\leq\,
		\frac{1}{4\sigma\ell^{1/12}}\,.$$
		On the other hand,
\begin{multline*}
	\lambda>
\sigma \Big(1-\displaystyle \frac{\kappa q}{\kappa-1}\Big)^{
	{(\kappa-1)\ell}/{\kappa}-\ell^{2/3}}\Big(1+\frac{1}{\ell^{1/13}}\Big)
	\qquad\Longrightarrow\qquad\cr
	\frac{1}{ \displaystyle \frac{\lambda}{
		\rho^{(\kappa-1)\ell/\kappa-\ell^{2/3}} }-1}\,\leq\,
	\frac{1}{ \displaystyle 
\sigma 
	\Big(1+\frac{1}{\ell^{1/13}}\Big)-1}\,.
\end{multline*}
	Applying the mean value theorem to the function $x\mapsto 1/(\sigma-1+x)$ on the interval
	$[0,\sigma/ {\ell^{1/13}}]$, we have
\begin{equation}
\label{bjor}
	\frac{1}{ \displaystyle \sigma 
	\Big(1+\frac{1}{\ell^{1/13}}\Big)-1}\,\leq\,
	\frac{1}{ \displaystyle \sigma -1}-
	\frac{1}{ (\displaystyle 2\sigma -1)^2}
	\frac{\sigma} {\ell^{1/13}}\,\leq\,
	\frac{1}{ \displaystyle \sigma -1}-
	\frac{1} {4\sigma\ell^{1/13}}
	\,.
\end{equation}
Combining the previous inequalities, we obtain that
\begin{equation*}
	\frac{\kappa-1}{\kappa^2\ell^{1/3}(\lambda-1)}\,+\,
	\frac{1}{ \displaystyle \frac{\lambda}{
		\rho^{(\kappa-1)\ell/\kappa-\ell^{2/3}} }-1}\,\leq\,
	\frac{1}{ \displaystyle \sigma -1}+
	\frac{1} {4\sigma}\Big(\frac{1}{\ell^{1/12}}-
		\frac{1}{\ell^{1/13}}
		\Big)\,<\,
	\frac{1}{ \displaystyle \sigma -1}\,,
\end{equation*}
which stands in contradiction with the inequality~\eqref{kambda} with $\varepsilon={\ell}^{-1/3}$. 
\end{proof}
\subsection{Revisiting the error threshold}
\label{ert}
Because of its importance,
we shall give two proofs of theorem~\ref{ierrt}.
The first proof is a soft proof
based on a compactness argument, whose starting point is 
the initial quasispecies equation.
The second proof is a more technical proof,
whose starting point is
the equation~\eqref{nlambda}
discovered by
Bratus, Novozhilov and Semenov.
Naturally, the second proof is more informative, it yields a control on the speed of convergence
and it opens the way to performing an asymptotic expansion of~$\lambda$ with respect to $\ell$ and $q$.

From equation~\eqref{meanfit}, we can express $x(w^*)$ as
\begin{equation}
\label{xm}
x(w^*)\,=\,\frac{\lambda-1}{\sigma-1}\,.
\end{equation}
Therefore it is enough to study the asymptotic behavior of $\lambda$.
We shall prove that
\begin{equation}
\label{dsym}
\lambda
\quad\longrightarrow\quad \max\big(1, \,\sigma e^{-a}\big)\,.
\end{equation}
\begin{proof}[Soft proof]
In the long chain regime, we have
\begin{equation}
\label{regim}
\sigma \big(1-q\big)^\ell
\quad\longrightarrow\quad \sigma e^{-a}\,.
\end{equation}
The quasispecies equation~\eqref{qs} associated with $w^*$ reads
\begin{equation}
\label{rwqs}
\lambda x(w^*)
	\,=\,
\sigma x(w^*)M(w^*,w^*)\,+
	\,\sum_{v\in E\setminus \{w^*\}}x(v)M(v,w^*)\,.
\end{equation}
We know that $x(w^*)>0$, therefore we deduce from~\eqref{rwqs} that
\begin{equation}
\label{lblwb}
\lambda 
	\,\geq\,
\sigma M(w^*,w^*)
	\,=\,
\sigma (1-q)^\ell
	\,.
\end{equation}
Using the lower bound~\eqref{lblwb} 
and the fact that $\lambda\geq 1$, we see that
\begin{equation}
\label{lisym}
\liminf\,\,\lambda\,\geq\,
\max\big(1, \,\sigma e^{-a}\big)\,.
\end{equation}
Suppose that $\lambda$ does not converge towards
$\max\big(1, \,\sigma e^{-a}\big)$. In this case, there exist $\varepsilon>0$ and
	two sequences of parameters $(\ell_n)_{n\geq 0}$,
	$(q_n)_{n\geq 0}$ such that
\begin{equation*}
\ell_n\to+\infty\,,\qquad
q_n\to 0\,,\qquad
\ell_n q_n\to a\,,
\end{equation*}
\begin{equation}
\label{vfg}
	\forall n\geq 0\qquad
	\lambda(\ell_n,q_n)\,\geq\,
\max\big(1, \,\sigma e^{-a}\big)+\varepsilon\,.
\end{equation}
Using~\eqref{xm} and~\eqref{vfg}, we obtain a lower bound on $x(w^*)$:
\begin{equation}
\label{lbx}
x(w^*)\,\geq\,\frac{\varepsilon}{\sigma-1}\,.
\end{equation}
We bound from above the quasispecies equation~\eqref{rwqs} associated with $w^*$, as follows:
\begin{equation}
\label{wqs}
\lambda x(w^*)
	\,\leq\,
	\sigma x(w^*)(1-q)^\ell+
	\,\big(1-x(w^*)\big)q\,,
\end{equation}
where we have used the fact that
	$M(v,w^*)\leq q$ whenever $v\neq w^*$.
Using together~\eqref{lbx} and \eqref{wqs}, we obtain that
\begin{equation*}
	\forall n\geq 0\qquad
	\lambda(\ell_n,q_n)
	\,\leq\,
	\sigma (1-q_n)^{\ell_n}+
	\frac{\sigma-1}{\varepsilon}
	q_n\,.
\end{equation*}
	Sending $n$ to $\infty$, we see that this inequality is not coherent with inequality~\eqref{vfg}.
	Therefore it must be the case that
$\lambda$ converges towards
$\max\big(1, \,\sigma e^{-a}\big)$.
\end{proof}
\begin{proof}[Technical proof]
	This proof is entirely based on
	the equation~\eqref{nlambda} discovered by
Bratus, Novozhilov and Semenov,
see formulas~$5.5$ and $5.6$ in \cite{BSN1},
and its probabilistic interpretation~\eqref{nnlambda}.
Let us start with a simple interesting inequality on $\lambda$. The map
$$\phi:x\in [0,+\infty[\,\,\mapsto\,
	\frac{1}{
	\displaystyle
		\frac{\lambda}{\rho^{x}}-1}\,
$$
is convex (recall that we suppose that $\rho>0$). 
The equation~\eqref{nlambda} (or rather \eqref{nnlambda}) can be rewritten with the help of the function~$\phi$ as
\begin{equation}
\label{rglambda}
	\frac{1}{\sigma-1}
	\,=\, \EE\left(\phi( {S_\ell}) \right)\,,
\end{equation}
where $S_\ell$ is a random variable with distribution the binomial law with parameters $\ell$ and $(\kappa-1)/\kappa$.
By the classical Jensen inequality, we have therefore
\begin{equation}
\label{jenslambda}
	\frac{1}{\sigma-1}
	\,\geq\, 
	\phi\big(\EE( {S_\ell})\big) 
	\,=\, 
 \phi\Big(\frac{\kappa-1}{\kappa}\ell\Big)
	\,.
\end{equation}
This yields the inequality
\begin{equation}
\label{minor}
\lambda\,\geq\,
	\sigma
	\rho^{ {(\kappa-1)\ell}/{\kappa}}
	\,=\,
	\sigma
	\Big(1-\displaystyle \frac{\kappa q}{\kappa-1}\Big)^{ {(\kappa-1)\ell}/{\kappa}
		}\,.
\end{equation}
	Notice that this inequality is not better than the 
lower bound~\eqref{lblwb} used in the soft proof.
	However it can certainly be improved with additional work.
In the long chain regime, we have
\begin{equation}
\label{tegim}
\sigma \Big(1-\displaystyle \frac{\kappa q}{\kappa-1}\Big)^{
	{(\kappa-1)\ell}/{\kappa}}
\quad\longrightarrow\quad \sigma e^{-a}\,.
\end{equation}
We put together the lower bound~\eqref{minor}, the upper bound~\eqref{major}
	(which have both been derived from the equation~\eqref{nlambda})
and the limit stated in~\eqref{tegim} to obtain the desired conclusion.
\end{proof}
\begin{figure}
\centering
\hspace*{-1 em}
\includegraphics[trim=0.6cm 0.5cm 0.4cm 1.1cm, clip=true, scale=0.63]{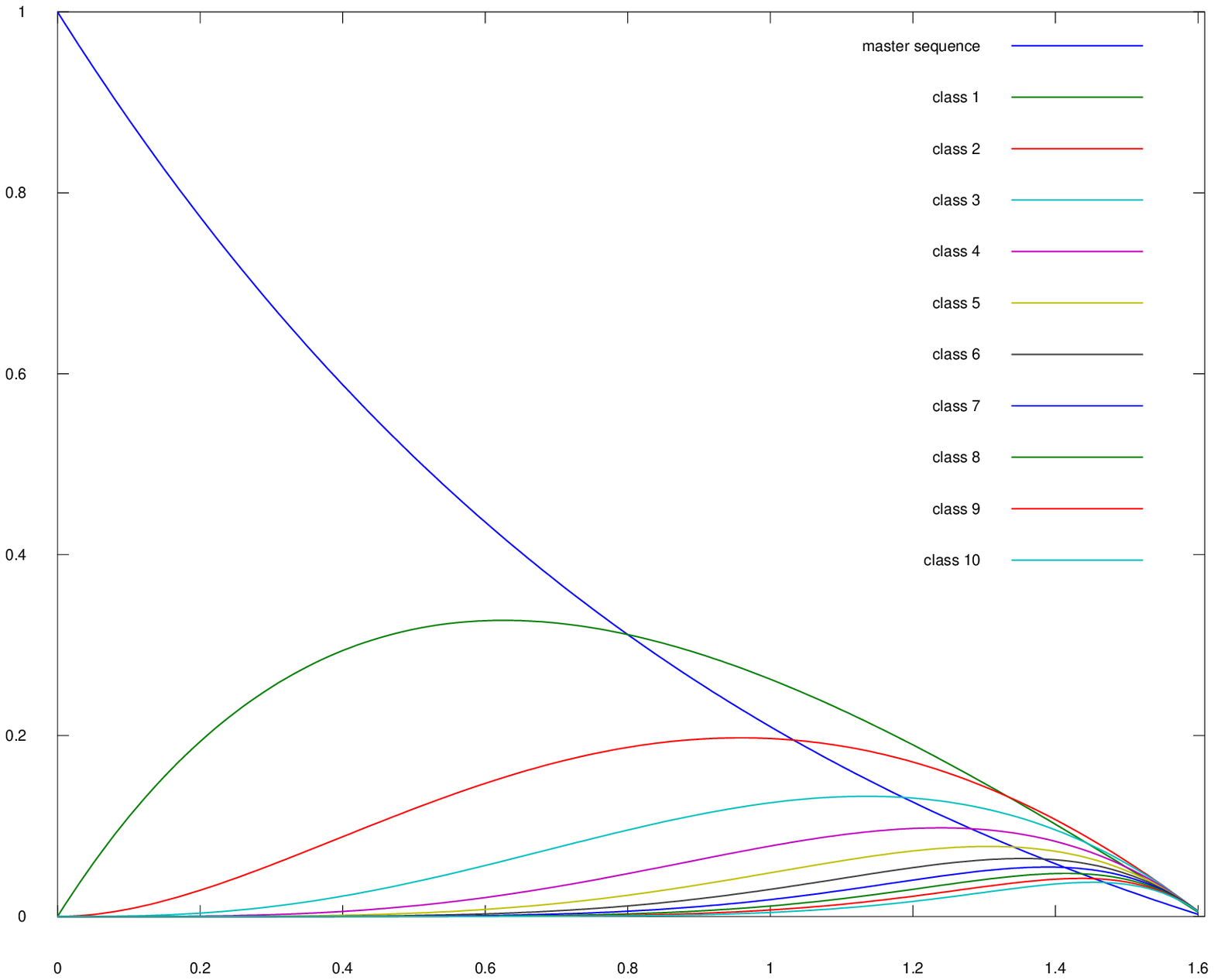}
\vspace{-9 pt}
\caption{
Fractions of the different types
as a function of $a$
for $\s=5$.}\label{f1}
\bigskip
\bigskip
%
%
 \hspace*{-1 em}
 \includegraphics[trim=0.6cm 0.5cm 0.4cm 0cm, clip=true, scale=0.63]{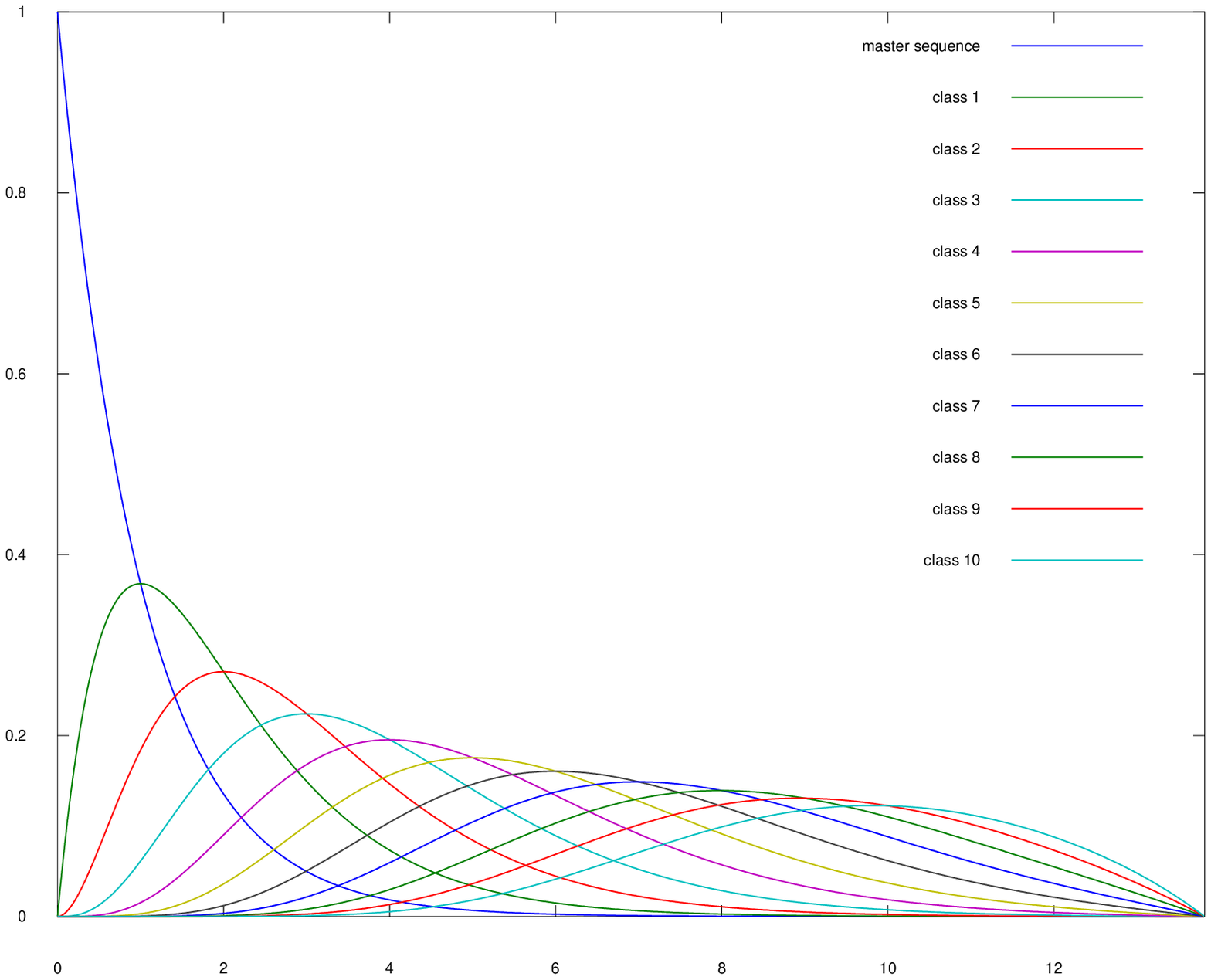}
 \vspace{-9 pt}
\caption{
Fractions of the different types
as a function of $a$
for $\s=10^6$.
}\label{f2}
\end{figure}
\noindent
\subsection{The limiting quasispecies distribution}
\label{lqd}
In section~\ref{exqs}, we computed an exact expression for the mean Hamming distance under the quasispecies distribution, which was
\begin{equation}
\label{rmeanHamm}
	\overline{\cQ} \,=\, 
	\sum_{k=0}^\ell k \,{\cQ(\s,a)}(k) \,=\, 
\frac{\ell q\, \lambda}{\lambda-\rho}\,.
\end{equation}
This formula is particularly illu\-mi\-na\-ting for the error threshold phe\-no\-me\-non.
We consider again
the long chain regime
\begin{equation}
\label{lcrb}
\ell\to+\infty\,,\qquad
q\to 0\,,\qquad
\ell q\to a\in\,[0,+\infty]\,.
\end{equation}
We have the dichotomy:

\noindent
$\bullet$ If $a>\ln \sigma$, then
$\lambda \rightarrow 1$, 
$\lambda-\rho\to 0$, whence
	$\overline{\cQ}
	\rightarrow +\infty$.

\noindent
$\bullet$ If $a<\ln \sigma$, then
$\lambda
\rightarrow \sigma e^{-a}>1$, and
\begin{equation}
\label{bamd}
	\overline{\cQ}
	\quad\longrightarrow\quad\frac{a\sigma\exa}{\sigma\exa-1}\,.
\end{equation}
We can go even further, indeed we can 
perform the expansion of formula~\eqref{probaclassek}
in the regime~\eqref{lcrb} and we obtain
$$
	\PP\big(X_{n}\in\cH_k\,\big|\,X_0=w^*\big)
	\,\sim\,
	\frac{\ell^k}{k!}(nq)^k(1-nq)^\ell
	\,\sim\,
	\frac{(na)^k}{k!}e^{-na}\,,
	$$
thus the asymptotic distribution of $X_n$ is the Poisson
distribution with parameter $na$.
Together with formula~\eqref{dsym}, this yields
that, for any fixed $n\geq 1$,
$$
\frac{1}{\lambda^n} 
	\PP\big(X_{n}\in\cH_k\,\big|\,X_0=w^*\big)
	\,\sim\,
	\frac{(na)^k}{\sigma^nk!}\,.
	$$
Substituting brutally this expansion into formula~\eqref{crobsol}, and using 
again for\-mu\-la~\eqref{dsym},
we get finally
\label{quas}
$$
	{\cQ}(k)
	\,\sim\,(\sigma e^{-a}-1)\frac{a^k}{k!}
\sum_{n\geq 1}\frac{n^k}{\sigma^n}\,.$$
Some steps in this computation must be made rigorous, 
but this can be done with a routine application of the dominated
convergence theorem.
In fact, this formula was obtained in \cite{CEDA}, while studying the quasispecies arising
in a stochastic model for the evolution of a finite population.
Thanks to this formula, we can easily draw the quasispecies distribution (figures~\ref{f1} and~\ref{f2}) and study
its dependence on the parameters $\sigma,a$. In the quasispecies literature, this
was previously done by integrating numerically the differential system (see for instance
\cite{doschu}, figure 3, p.9).
Besides, we can now perform simple theoretical computations with the help of this distribution.
For instance, 
the mutation rate at which the proportion of the Hamming class $1$ becomes larger than the proportion of the
wild type $w^*$ is
$a=1-1/{\sigma}$.

\section{Extension to a general landscape}
\label{ext}
\noindent
Our starting point is the quasispecies equation~\eqref{qs}, rewritten in the following form:
\begin{equation}
\label{rqs}
\forall\, u\in E\qquad
\lambda x(u) \,=\,
\,\sum_{v\in E}x(v)f(v)M(v,u)\,
\end{equation}
subject to the constraint
\begin{equation}
\label{rcons}
\forall\, u\in E\qquad x(u)\geq 0\,,\qquad
\sum_{u\in E}x(u)\,=\,1\,,  
\end{equation}
where
	$$\lambda\,=\,\sum_{v\in E}x(v)f(v)$$
	is the mean fitness, or equivalently the Perron-Frobenius eigenvalue of the matrix
	$(f(v)M(v,u))_{u,v\in E}$.
\subsection{Reformulation of the quasispecies equations}
The computations performed in this subsection are in fact valid without
any assumption on the geometry of the genotype space~$E$.
	Thanks to the hypothesis on the fitness function, we have that $\lambda\geq 1$.
Moreover, since the matrix $M$ has only positive entries
	 and since the fitness function is not
	identically equal to~$1$, then the Perron-Frobenius eigenvalue is strictly larger than one,
	i.e., $\lambda>1$.
We rewrite the equation~\eqref{rqs} as
\begin{equation}
\label{rrqs}
\forall\, u\in E\qquad
\lambda x(u) 
-\,\sum_{v\in E}x(v)M(v,u)\,
	\,=\,
	\,\sum_{v\in E}x(v)\big(f(v)-1\big)M(v,u)\,.
\end{equation}
We denote by $x^t$ the row vector which is the transpose of the column
vector~$x$ and we introduce the square matrix whose diagonal coefficients are the fitness values $(f(u),u\in E)$:
\begin{equation}
    \label{eigenfactorise}
F\,=\,
\begin{pmatrix}
	\ddots &  0 & 0\cr
	0 & f(u) & 0\cr
	0 & 0 & \ddots\cr
\end{pmatrix}\,.
\end{equation}
After dividing by $\lambda$ (recall that $\lambda> 1$), equation~\eqref{rrqs} can be rewritten in
matrix form as
\begin{equation}
    \label{mform}
x^t\Big(I-\frac{1}{\lambda}M\Big)\,=\,
x^t\frac{1}{\lambda}\big(F-I\big)M\,.
\end{equation}
Now the matrix $M$ is a stochastic matrix, hence its spectral radius is equal to $1$. Since $\lambda>1$, then
the matrix
$I-\frac{1}{\lambda}M$ is invertible, and its inverse is given by the geometric series:
$$ \Big(\text{I}-\frac{M}{\lambda}\Big)^{-1}\, =\,\, \sum_{n\geq 0} \frac{M^n}{\lambda^n}\,.
    $$
Plugging this identity into equation~\eqref{mform}, we obtain
\begin{equation}
    \label{eigendvpe}
x^t\,=\,
x^t
	\big(F-I\big)
    \sum_{n\geq 1} \frac{M^n}{\lambda^n}\,.
\end{equation}
Let us introduce the set $W^*$ of the genotypes where the fitness function is strictly larger than one:
$$W^*\,=\,\big\{\,u\in E:f(u)>1\,\big\}\,.$$
We call the set $W^*$ the set of the wild types.
We introduce the matrix $N_\lambda$, indexed by the elements of $W^*$, defined as
\begin{equation}
    \label{nma}
\forall v,w\in W^*\qquad
N_\lambda(v,w)\,=\,
	\Big(\big(F-I\big)
    \sum_{n\geq 1} \frac{M^n}{\lambda^n}\Big)(v,w)\,.
\end{equation}
With this notation, the linear system 
    \eqref{eigendvpe}
of size $\kappa^\ell$ can be split into the system of size $|W^*|$ given by
\begin{equation}
    \label{gendvpe}
\forall v\in W^*\qquad
x(v)\,=\,\sum_{w\in W^*}
x(w)\,
N_\lambda(w,v)\,,
\end{equation}
and the
$\kappa^\ell-|W^*|$ remaining equations
\begin{equation}
    \label{rendvpe}
\forall v\in E\setminus W^*\qquad
x(v)\,=\,\sum_{w\in W^*}
x(w)(f(w)-1)\,
    \sum_{n\geq 1} \frac{M^n}{\lambda^n}(w,v)\,.
\end{equation}
The system~\eqref{gendvpe} expresses that the vector 
$(x(v),\,v\in W^*)$ is a left Perron-Frobenius eigenvector of the matrix $N_\lambda$,
and that the Perron-Frobenius eigenvalue of $N_\lambda$ is equal to $1$.
The equations~\eqref{rendvpe} show that the remaining coordinates 
$(x(v),\,v\in E\setminus W^*)$ are completely determined by 
$(x(v),\,v\in W^*)$. 
These considerations suggest that the quasispecies equation can in principle be solved through
the following procedure.

\noindent
{\bf General strategy.}
We start by isolating the set $W^*$ of the wild types and we form the matrix
$(N_\lambda(w,v))_{v,w\in W^*}$, defined in formula~\eqref{nma}, where $\lambda> 1$ is 
considered as a parameter.
Now, the Perron-Frobenius eigenvalue of 
$N_\lambda$ is a decreasing function of $\lambda$, which tends to $0$ as $\lambda$ goes to $+\infty$
and to $+\infty$ as $\lambda$ goes to $1$.
We choose for $\lambda$ the unique value such that the Perron-Frobenius eigenvalue of $N_\lambda$ is equal to $1$.
Once this value is fixed, we solve the system \eqref{gendvpe} on~$W^*$.
The remaining coordinates of $x$ on $E\setminus W^*$ are determined by 
equations~\eqref{rendvpe}.

\noindent
{\bf A general lower bound on~$\lambda$.}
We close this section with a general lower bound on~$\lambda$.
The quasispecies equations imply that
$$\forall u\in E\qquad\lambda x(u)\,\geq\,x(u)\,f(u)\,M(u,u)\,=\,x(u)f(u)\,(1-q)^\ell\,.$$
Since $x(u)>0$ for any $u\in E$, this readily implies that
\begin{equation}
    \label{lwb}
	\lambda \,\geq\,
	\big(\max_Ef\big)\,(1-q)^\ell\,.
\end{equation}
\subsection{Computation of 
$ \sum_{n\geq 1} {M^n}/{\lambda^n}$}
\label{commn}
From now onwards,
we work with the genotype space $E=\cA^\ell$
where $\cA$ is a finite alphabet of cardinality $\kappa$ 
and we consider the mutation scheme~\eqref{mut}. 
We start by computing $M^n$. We use the notations and the technique of section~\ref{mutdyn} and
we generalize formula~\eqref{comput} as follows:
\begin{align}\label{rcomput}
	M^n(u,v)&\,=\,
	\PP\big(X_{n}=v\,\big|\,X_0=u\big)\,=\,
\PP\big(X_{n}(i)=v(i), \,1\leq i\leq\ell\,\big|\,X_0=u\big)\cr
		&\,=\,
	{\prod_{1\leq i\leq\ell}} 
	\PP\big(X_{n}(i)=v(i)\,\big|\,X_0(i)=u(i)\big)\,.
\end{align}
Now, we have
\begin{equation}\label{romput}
	\PP\big(X_{n}(i)=v(i)\,\big|\,X_0(i)=u(i)\big)\,=\,
	\begin{cases}
		\displaystyle \frac{1}{\kappa}
		+ 
\displaystyle \frac{\kappa-1}{\kappa}\rho^n  
		\phantom{\frac{1}{\Big|}} 
		& \text{if }u(i)=v(i)\,,\cr 
\displaystyle \frac{1}{\kappa}\big(1-\rho^n \big)& 
		 \text{if }u(i)\neq v(i)\,,\cr 
	\end{cases}
\end{equation}
where 
$$\rho\,= \,1-\frac{\kappa q}{\kappa-1}\,.$$
Coming back to formula~\eqref{rcomput}, we conclude that
\begin{equation}\label{ucomput}
	M^n(u,v)\,=\,
	\displaystyle \Big(\frac{1}{\kappa} + 
	\displaystyle \frac{\kappa-1}{\kappa}\rho^n\Big)^{\ell-d(u,v)}  
	\displaystyle \Big(\frac{1}{\kappa}\big(1-\rho^n\big) \Big)^{d(u,v)}\,.
\end{equation}
We wish now to compute 
$$  \sum_{n\geq 1} \frac{M^n}{\lambda^n}\,.  $$
Our first strategy consists of developing formula~\eqref{ucomput}.
We end up with several geometric series
and we obtain a closed finite formula. So, starting from~\eqref{ucomput}, we have, setting $d=d(u,v)$ to alleviate
the notation,
\begin{equation}\label{bcomput}
	M^n(u,v)\,=\,
	\sum_{i=0}^{\ell-d}
	\sum_{j=0}^{d}
	\binom{\ell-d}{i}
	\displaystyle \Big(\frac{1}{\kappa}\Big)^{\ell-d-i}  
	\displaystyle \Big(\frac{\kappa-1}{\kappa}\Big)^i\rho^{ni}
	\binom{d}{j}
	\displaystyle \frac{1}{\kappa^d}
	(-1)^j
	\rho^{nj}\,.
\end{equation}
We can now sum the geometric series and we get
\begin{equation}\label{vcomput}
	\sum_{n\geq 1} \frac{M^n}{\lambda^n}(u,v)\,=\,
	\sum_{i=0}^{\ell-d}
	\sum_{j=0}^{d}
	\binom{\ell-d}{i}
	\displaystyle \Big(\frac{1}{\kappa}\Big)^{\ell-d-i}  
	\displaystyle \Big(\frac{\kappa-1}{\kappa}\Big)^i
	\binom{d}{j}
	\displaystyle \frac{1}{\kappa^d}
	\displaystyle \frac{
	(-1)^j
		}{
		\displaystyle \frac{\lambda}{\rho^{i+j}}-1
		}
	\,.  
\end{equation}
\subsection{Probabilistic interpretation}
We provide furthermore a probabilistic interpretation of the apparently complex formula~\eqref{vcomput}.
Let $S_{\ell-d}$ be a binomial random variable with parameters $\ell-d,1-1/\kappa$. We have
\begin{equation}\label{vpcomput}
	\sum_{n\geq 1} \frac{M^n}{\lambda^n}(u,v)\,=\,
	\EE\Bigg(
	\frac{1}{\kappa^d}
	\sum_{j=0}^{d}
	\binom{d}{j}
	\displaystyle 
	\displaystyle \frac{
	(-1)^j
		}{
			\displaystyle \frac{\lambda}{\rho^{S_{\ell-d}+j}}-1}
		\Bigg)
	\,.  
\end{equation}
The above formula is quite nice. Its drawback is that it contains negative terms, so it is not obvious
to see that the global result is non-negative. 
We present next a third formula, which avoids this problem.
\subsection{Third formula}
The trick consists of developing the factor $(1-\rho^n)^d$ of formula~\eqref{ucomput}
in a more complicated way, as follows:
\begin{equation}\label{icomput}
	M^n(u,v)\,=\,
	\sum_{i=0}^{\ell-d}
	\binom{\ell-d}{i}
	\displaystyle \Big(\frac{1}{\kappa}\Big)^{\ell-d-i}  
	\displaystyle \Big(\frac{\kappa-1}{\kappa}\Big)^i
	\rho^{ni}
	\displaystyle \frac{1}{\kappa^d}
	(1-\rho)^d
	\Big(\sum_{j=0}^{n-1}\rho^j\Big)^{d}
	\,.
\end{equation}
We wish to compute the series $\sum_{n\geq 1} {M^n}/{\lambda^n}$. 
From now onwards, we deal with the case $d\geq 1$ (for $d=0$, we use the expression
obtained in~\eqref{vcomput}, where all the terms are positive).
Focusing on the two terms 
which depend on $n$ in formula~\eqref{icomput}, we end up with the series
\begin{align}\label{tract}
	\sum_{n\geq 1} \frac{ \rho^{ni} }{\lambda^n}
	\Bigg(\sum_{j=0}^{n-1}\rho^j\Bigg)^{d}
	&\,=\,
	\sum_{n\geq 1} \frac{ \rho^{ni} }{\lambda^n}
	\sum_{j_1=0}^{n-1} \cdots
	\sum_{j_d=0}^{n-1}
	\rho^{j_1+\cdots+j_d}\cr
	&\,=\,
	\sum_{j_1,\dots,j_d\geq 0}\,\,
	\sum_{n> \max(j_1,\dots,j_d)}
	\frac{ \rho^{ni} }{\lambda^n}
	\rho^{j_1+\cdots+j_d}\cr
	&\,=\,
	\sum_{j_1,\dots,j_d\geq 0}\,\,
	\frac{
		\displaystyle{\Big(\frac{ \rho^{i} }{\lambda}\Big)^{
			\max(j_1,\dots,j_d)+1}\kern-11pt
			}
		}{
			\displaystyle 1-\frac{ \rho^{i} }{\lambda}}
	\,\,\rho^{j_1+\cdots+j_d}
	\,.
\end{align}
Notice that this formula is not quite legitimate for $d=0$, unless we make the convention
that, for $d=0$, there is only one term in the sum corresponding to $j_1=\cdots=j_d=0$. Therefore 
it seems safer to perform the above computation only when $d\geq 1$.
Let us set $\gamma= { \rho^{i} }/{\lambda}$ and let us focus on the sum
\begin{equation}\label{focus}
	\sum_{j_1,\dots,j_d\geq 0}\,\,
		\displaystyle{\gamma^{
			\max(j_1,\dots,j_d)}
			}
	\,\,\rho^{j_1+\cdots+j_d}
	\,.
\end{equation}
We shall decompose the sum according to the number~$r$ of distinct indices among $j_1,\cdots,j_d$, their different values
$i_1,\dots,i_r$,
and the number of times $n_1,\dots,n_r$ each value appears. We get
\begin{equation}\label{decomp}
	\sum_{r=1}^d
	\sum_{0\leq i_1<\cdots< i_r}\,\,
	\sum_{\genfrac{}{}{0pt}{1}{n_1,\dots,n_r\geq 1}{n_1+\cdots+n_r=d}}
	\sum_{j_1,\dots,j_d\in\mathcal J}
		\displaystyle{\gamma^{ \max(j_1,\dots,j_d)} }
	\,\,\rho^{j_1+\cdots+j_d}
	\,,
\end{equation}
where the innermost sum extends over the set $\mathcal J$ defined
by
\begin{equation*}
	{\mathcal J}\,=\,
	\Big\{\,j_1,\dots,j_d\geq 0:
	{\text{ for }}1\leq k\leq d,
	\,i_k\text{ appears }{n_k\text{ times in }} {j_1,\dots,j_d}
\,\Big\} \,.
\end{equation*}
The number of terms appearing in the last summation corresponds to the number of placements of $d$ balls in
$r$ cells with occupancy numbers given by $n_1,\dots,n_r$ (see for instance the classical book of Feller \cite{FE}, section~II.5),
that is
$d!/(n_1!\cdots n_r!)$. For each of these placements, we have
		$${ \max(j_1,\dots,j_d)} \,=\,i_r\,,\qquad
	{j_1+\cdots+j_d}\,=\,n_1i_1+\cdots +n_ri_r\,,$$
	therefore the formula~\eqref{decomp} becomes
\begin{equation}\label{lecomp}
	\sum_{r=1}^d
	\sum_{0\leq i_1<\cdots< i_r}\,\,
	\sum_{\genfrac{}{}{0pt}{1}{n_1,\dots,n_r\geq 1}{n_1+\cdots+n_r=d}}
	\frac{d!}{n_1!\cdots n_r!}
		\gamma^{ i_r} 
	\,\,\rho^{ n_1i_1+\cdots +n_ri_r }
	\,.
\end{equation}
We can perform the summation over $i_r$, and we get
\begin{multline}\label{eecomp}
	\sum_{0\leq i_1<\cdots< i_r}\,\,
		\gamma^{ i_r} 
	\,\,\rho^{ n_1i_1+\cdots +n_ri_r }\,=\,\cr
		\frac{\gamma\rho^{n_r}} {1-\gamma\rho^{n_r}}
	\sum_{0\leq i_1<\cdots< i_{r-1}}\kern-3pt
		\big(\gamma\rho^{n_r}\big)^{ i_{r-1}} 
	\,\,\rho^{
		n_1i_1+\cdots +n_{r-1}i_{r-1}
	}
	\,.
\end{multline}
The number of indices $i_1,\dots,i_{r-1}$ has decreased by $1$, and $\gamma$ has been replaced by
		$\gamma\rho^{n_r}$. We proceed in the same
		way and we perform successively the summations over
		$i_{r-1},\dots,i_1$ until we obtain
\begin{multline*}
	\sum_{0\leq i_1<\cdots< i_r}\,\,
		\gamma^{ i_r} 
	\,\,\rho^{ n_1i_1+\cdots +n_ri_r }\,=\,\cr
		\frac{\gamma\rho^{n_r}} {1-\gamma\rho^{n_r}}
		\frac{\gamma\rho^{n_r+n_{r-1}}} {1-\gamma\rho^{n_r+n_{r-1}}}
		\cdots
		\frac{\gamma\rho^{n_r+\cdots+n_{2}}} {1-\gamma\rho^{n_r+\cdots+n_{2}}}
	\sum_{0\leq i_1}\,\,
		\big(\gamma
	\rho^{ n_r+\cdots +n_2 }\big)^{i_1}
\rho^{ n_1i_1}\,.
\end{multline*}
Recalling that $n_1+\cdots+n_r=d$, we can compute the last sum
	$$\sum_{0\leq i_1}\,\,
		\big(\gamma
	\rho^{ n_r+\cdots +n_2 }\big)^{i_1}
\rho^{ n_1i_1}
\,=\,
		\frac{1} {1-\gamma\rho^{d}}
$$
and we conclude that
\begin{multline}\label{vcomp}
	\sum_{0\leq i_1<\cdots< i_r}\,\,
		\gamma^{ i_r} 
	\,\,\rho^{ n_1i_1+\cdots +n_ri_r }\,=\,\cr
		\frac{\gamma\rho^{n_r}} {1-\gamma\rho^{n_r}}
		\frac{\gamma\rho^{n_r+n_{r-1}}} {1-\gamma\rho^{n_r+n_{r-1}}}
		\cdots
		\frac{\gamma\rho^{n_r+\cdots+n_{2}}} {1-\gamma\rho^{n_r+\cdots+n_{2}}}
		\frac{1} {1-\gamma\rho^{d}}
	\,.
\end{multline}
Putting together
formulas~\eqref{lecomp} and~\eqref{vcomp},
we obtain a finite formula for the sum~\eqref{focus}:
\begin{multline}\label{ficomp}
	\sum_{j_1,\dots,j_d\geq 0}\,\,
		\displaystyle{\gamma^{
			\max(j_1,\dots,j_d)}
			}
	\,\,\rho^{j_1+\cdots+j_d}
	\,=\,\cr
	\sum_{r=1}^d
	\sum_{\genfrac{}{}{0pt}{1}{n_1,\dots,n_r\geq 1}{n_1+\cdots+n_r=d}}
	\frac{d!}{n_1!\cdots n_r!}
	\frac{1}{\gamma\rho^{d}}
		\frac{\rho^{rn_r+(r-1)n_{r-1}+\cdots+n_{1}}}
			{\prod_{k=1}^r \Big({\frac{1}{\gamma}-\rho^{n_r+\cdots+n_{r-k+1}}}\Big)}
	\,.
\end{multline}
Plugging this formula into~\eqref{tract} and \eqref{icomput}
yields that, for $d=d(u,v)\geq 1$, 
\begin{multline}\label{wcomput}
	\sum_{n\geq 1} \frac{M^n}{\lambda^n}(u,v)\,=\,
	\sum_{i=0}^{\ell-d}
	\binom{\ell-d}{i}
	\displaystyle \Big(\frac{1}{\kappa}\Big)^{\ell-i}  
	\displaystyle \Big(\frac{\kappa-1}{\kappa}\Big)^i
	\displaystyle \Big(\frac{1-\rho}{\rho}\Big)^d
	\frac{ 1 }{1-\frac{\rho^i}{\lambda}}\cr
\times	\sum_{r=1}^d
	\sum_{\genfrac{}{}{0pt}{1}{n_1,\dots,n_r\geq 1}{n_1+\cdots+n_r=d}}
	\frac{d!}{n_1!\cdots n_r!}
		\frac{\rho^{rn_r+(r-1)n_{r-1}+\cdots+n_{1}}}
			{\prod_{k=1}^r \Big({\frac{\lambda}{\rho^i}-\rho^{n_r+\cdots+n_{r-k+1}}}\Big)}
	\,.
\end{multline}
We can further rewrite this formula using the expectation with respect to a Binomial random variable
$S_{\ell-d}$ with parameters $\ell-d$ and $1-1/\kappa$. Indeed, we have
\begin{multline}
	\label{xcomput}
	\sum_{n\geq 1} \frac{M^n}{\lambda^n}(u,v)\,=\,
	\displaystyle \frac{1}{\kappa^d}
	\displaystyle \Big(\frac{1-\rho}{\rho}\Big)^d
\times	\cr
	\EE\Bigg(
	\frac{ 1 }{1-\frac{\rho^{S_{\ell-d}}}{\lambda}}
	\sum_{r=1}^d
	\sum_{\genfrac{}{}{0pt}{1}{n_1,\dots,n_r\geq 1}{n_1+\cdots+n_r=d}}
	\frac{d!}{n_1!\cdots n_r!}
		\frac{\rho^{rn_r+(r-1)n_{r-1}+\cdots+n_{1}}}{\prod_{k=1}^r 
		\Big(\frac{\lambda}{\rho^{S_{\ell-d}}}-\rho^{n_r+\cdots+n_{r-k+1}}\Big)}
			\Bigg)\,.
\end{multline}
Each of the previous formulas has its own interest and might be useful, depending on the context.
For instance, with formula~\eqref{vcomput}, we see directly that the coefficients of the matrix
$N_\lambda$ are rational functions of the parameter $\lambda$, and that there exists a unique choice 
for $\lambda$ such that the Perron-Frobenius eigenvalue of $N_\lambda$ becomes equal to $1$.
The analysis of the error threshold conducted in section~\ref{ert} rests entirely on the probabilistic
representation presented in formula~\eqref{vpcomput} in the specific case where $u=v=w^*$ and $d=0$.
Finally, the third formula~\eqref{xcomput} will be useful to analyze the asymptotic behavior of the
non-diagonal entries of the matrix $N_\lambda$ for the landscape with finitely many peaks
in section~\ref{fmape}.
\subsection{Proof of 
theorem~\ref{relationFH}}
Let $G:E\to\R$ be an additive functional, that is a function given by
$$\forall u\in E\qquad G(u)\,=\,\sum_{i=1}^\ell g(u(i))\,,$$
where $g$ is a function defined on $\cA$
with values in $\R$.
We consider an arbitrary fitness function satisfying hypothesis~\ref{exhyp},
and we denote by $(x(u), u\in E)$, the solution of the quasispecies equation associated with $f$.
Our goal here is to relate the mean value $\oG$ of $G$ in the quasispecies 
$(x(u), u\in E)$, defined by
$$\oG\,=\,\sum_{u\in E}G(u)x(u)\,$$
to the mean fitness $\lambda$ or $\of$, defined by
$$\of\,=\,\sum_{u\in E}f(u)x(u)\,.$$
Our starting point is formula~\eqref{eigendvpe}, which yields
\begin{align}\label{stpo}
	\oG &\,=\,
\sum_{u\in E}G(u)
\sum_{v\in E} x(v)
	\big(f(v)-1\big)
    \sum_{n\geq 1} \frac{1}{\lambda^n}M^n(v,u)
    \cr
	&\,=\,
\sum_{v\in E} x(v)
	\big(f(v)-1\big)
    \sum_{n\geq 1} 
    \frac{1}{\lambda^n}
\sum_{u\in E} M^n(v,u)\, G(u)
    \,.
\end{align}
We compute next the most inner sum. By the definition of $G$,
\begin{equation}
    \label{defog}
\sum_{u\in E} M^n(v,u)\, G(u)\,=\,
	\sum_{i=1}^\ell\sum_{u\in E} M^n(v,u)\, g(u(i))\,.
\end{equation}
Moreover, formulas~\eqref{rcomput} and \eqref{romput} yield that, for $1\leq i\leq\ell$, 
\begin{align}\label{mois}
	\sum_{u\in E} M^n(v,u)\, g(u(i))&\,=\,
		\Big(\displaystyle \frac{1}{\kappa} + \displaystyle \frac{\kappa-1}{\kappa}\rho^n  \Big)g(v(i))
		+
	\Big(\displaystyle \frac{1}{\kappa}\big(1-\rho^n \big) \Big)\sum_{a\in\cA\setminus\{v(i)\}}g(a)
	\cr
					&\,=\, \rho^n g(v(i)) + \displaystyle \big(1-\rho^n \big) \bg\,,\kern20pt
\end{align}
where  $\bg$ is the mean value of $g$ over the alphabet $\cA$, i.e.,
\begin{equation}
    \label{bbg}
    \bg\,=\,
	\frac{1}{\kappa}
	\sum_{a\in\cA}g(a)
	\,.
\end{equation}
Plugging formula~\eqref{mois} in formula~\eqref{defog}, we obtain
\begin{equation}
    \label{refog}
\sum_{u\in E} M^n(v,u)\, G(u)\,=\,
	 \rho^n G(v) + \displaystyle \ell\big(1-\rho^n \big) \bg\,.
\end{equation}
Inserting this last formula into~\eqref{stpo}, we obtain
\begin{align}\label{sjpo}
	\oG &\,=\,
\sum_{v\in E} x(v)
	\big(f(v)-1\big)
	\sum_{n\geq 1} \frac{1}{\lambda^n}\Big(
	 \rho^n G(v) + \displaystyle \ell\big(1-\rho^n \big) \bg\Big)
    \cr
	&\,=\,
\sum_{v\in E} x(v)
	\big(f(v)-1\big)
	\Big(
	 \frac{\rho}{\lambda-\rho} G(v) + \displaystyle \frac{\ell\bg}{\lambda-1}-
	 \frac{\ell\bg\rho}{\lambda-\rho} 
	 \Big)
    \cr
	&\,=\,
	 \frac{\rho}{\lambda-\rho}
	 \big( \ofG-\oG\big) + 
	 \big( \of-1\big)  
	 \displaystyle {\ell\bg}
	 \frac{\lambda(1-\rho)}{(\lambda-1)(\lambda-\rho)} 
    \,,
\end{align}
where we have introduced the notation
\begin{equation}
    \label{fbg}
    \ofG
	\,=\,\sum_{u\in E}f(u)\,G(u)\,x(u)\,.
\end{equation}
Recalling that $\of=\lambda$, we have proved the curious formula stated in the theorem~\ref{relationFH}.
\section{Finitely many peaks}
\label{fmape}
As for the sharp peak landscape,
we shall give two proofs of theorem~\ref{ifrrt}, a soft one, whose starting point is 
the initial quasispecies equation,
and a technical one, based on the strategy explained at the end of section~\ref{ext}.
Again, the soft proof might look easier, but the technical proof is more informative and could 
potentially yield more accurate estimates.
We define first
the set $W^*$ of the wild types as
$$W^*\,=\,\big\{\,w_i:1\leq i\leq k\,\big\}\,.$$
Throughout the proofs, we write simply $\lambda(\ell,q)$ or even $\lambda$ instead of
	$\lambda_{\text{FP}}(\ell,q)$.
\subsection{Soft proof}
	We start as in the case of the sharp peak landscape.
In the long chain regime, we have that
	$\sigma\smash{ \big(1-q\big)^\ell}\to
	\sigma e^{-a}$.
Using the
lower bound~\eqref{lwb} and the fact that $\lambda\geq 1$, we see that
\begin{equation}
\label{flisym}
\liminf\,\,\lambda\,\geq\,
\max\big(1, \,\sigma e^{-a}\big)\,.
\end{equation}
Suppose that $\lambda$ does not converge towards
$\max\big(1, \,\sigma e^{-a}\big)$. In this case, there exist $\varepsilon>0$ and
	two sequences of parameters $(\ell_n)_{n\geq 0}$,
	$(q_n)_{n\geq 0}$ such that
\begin{equation*}
\ell_n\to+\infty\,,\qquad
q_n\to 0\,,\qquad
\ell_n q_n\to a\,,
\end{equation*}
\begin{equation}
\label{zfg}
	\forall n\geq 0\qquad
	\lambda(\ell_n,q_n)\,\geq\,
\max\big(1, \,\sigma e^{-a}\big)+\varepsilon\,.
\end{equation}
For $1\leq i\leq k$,
we bound from above the quasispecies equation~\eqref{qs} associated with $w^*_i$, as follows:
\begin{equation}
\label{zqs}
\lambda x(w^*_i)
	\,\leq\,
	\sum_{v\in {W^*}}
\sigma x(v)M(v,w^*_i)\,+
	\,\sum_{v\in E\setminus {W^*}}x(v)M(v,w^*_i)\,.
\end{equation}
	We sum equation~\eqref{zqs} over $i\in\{\,1,\dots,k\,\}$.
	Setting
	$$\displaylines{
		x(W^*)\,=\,\sum_{1\leq i\leq k}x(w^*_i)\,,\cr
		\forall v\in E
	\qquad M(v,W^*)\,=\,\sum_{1\leq i\leq k}M(v,w^*_i)\,,}
	$$
	we obtain
\begin{equation}
\label{zqz}
	\lambda x(W^*)
	\,\leq\,
	\sum_{v\in {W^*}}
\sigma x(v)M(v,W^*)\,+
	\,\sum_{v\in E\setminus {W^*}}x(v)M(v,W^*)\,.
\end{equation}
We use next the fact that $M(v,w^*_i)\leq q$
	 whenever $v\neq w^*_i$ and we obtain that
	$$\forall v\in W^*\qquad M(v,W^*)\,\leq\,M(v,v)+kq\,=\,(1-q)^\ell +kq\,,$$
	$$\forall v\in E\setminus W^*\qquad M(v,W^*)\,\leq\,kq\,.$$
	Inserting these inequalities in~\eqref{zqz}, we obtain that
\begin{align}
\label{zzz}
	\lambda x(W^*)
	&\,\leq\,
	\sigma x(W^*)\big((1-q)^\ell+kq\big)+
	\,\big(1-x(W^*)\big)kq\cr
	&\,\leq\,
	\sigma x(W^*)(1-q)^\ell+\sigma kq
	\,.
\end{align}
In addition, we have that
\begin{equation}
\label{zpz}
	\lambda 
	\,\leq\, \sigma x(W^*)+1-x(W^*)
	\,\leq\, \sigma x(W^*)+1\,.
\end{equation}
Equation~\eqref{zpz} and the condition~\eqref{zfg} yield that
\begin{equation}
\label{vpz}
	\frac{1}{x(W^*)}
	\,\leq\,\frac{ \sigma}{\lambda-1}
	\,\leq\,\frac{ \sigma}{\varepsilon}\,. 
\end{equation}
Using together~\eqref{zzz} and \eqref{vpz}, we obtain that
\begin{equation*}
	\forall n\geq 0\qquad
	\lambda(\ell_n,q_n)
	\,\leq\,
	\sigma (1-q_n)^{\ell_n}+
	\frac{\sigma}{\varepsilon}
	\sigma k
	q_n\,.
\end{equation*}
	Sending $n$ to $\infty$, we see that this inequality is not coherent with inequality~\eqref{zfg}.
	Therefore it must be the case that
$\lambda$ converges towards
$\max\big(1, \,\sigma e^{-a}\big)$.
\subsection{Technical proof}
	For this proof, we shall implement
	the strategy explained at the end of section~\ref{ext}.
	So we consider the matrix $\big(N_\lambda(i,j)\big)_{1\leq i,j\leq k}$ defined by
\begin{equation}
    \label{fmpma}
	\forall i,j\in \{\,1,\dots,k\,\}\qquad
N_\lambda(i,j)\,=\,
	\Big(\big(F-I\big)
    \sum_{n\geq 1} \frac{M^n}{\lambda^n}\Big)(w^*_i,w^*_j)\,,
\end{equation}
where 
the matrix $F$ was introduced in formula~\eqref{eigenfactorise}.
The advantage is that the size of the matrix $N_\lambda$ is $k\times k$ and this size does not vary with $\ell$ and $q$.
The diagonal elements of the matrix are given by
$$N_\lambda(i,i)\,=\,
	(\sigma_i-1)
    \sum_{n\geq 1} \frac{1}{\lambda^n}M^n(w^*_i,w^*_i)\,,\quad 1\leq i\leq k\,.
    $$
In fact, the above series does not depend on~$i$. Using formula~\eqref{vpcomput} with $d=0$, we have
\begin{equation}
    \label{nij}
    \forall i\in\{\,1,\dots,k\,\}\qquad
N_\lambda(i,i)\,=\,
	(\sigma_i-1)\,
	\EE\Bigg(
	\displaystyle 
	\displaystyle \frac{
		1}{
			\displaystyle \frac{\lambda}{\rho^{S_{\ell}}}-1}
		\Bigg)
	\,,  
\end{equation}
where $S_{\ell}$ is a binomial random variable with parameters $\ell,1-1/\kappa$. 
The parameter $\lambda$ is adjusted so that the Perron-Frobenius eigenvalue of the matrix
$N_\lambda$ is equal to~$1$. This Perron-Frobenius eigenvalue is always larger or equal than
any diagonal element, thus we must have
	$N_\lambda(i,i)\,\leq \,1$. Proceeding exactly as in the technical proof of theorem~\ref{ierrt}, we obtain
	with the help of Jensen's inequality that
\begin{equation}
\label{rinor}
    \forall i\in\{\,1,\dots,k\,\}\qquad
\lambda\,\geq\,\sigma_i
	\rho^{
		{(\kappa-1)\ell}/{\kappa}
		}\,.
\end{equation}
Taking the maximum over $i\in\{\,1,\dots,k\,\}$, we get
\begin{equation}
\label{mrinor}
\lambda\,\geq\,\sigma
	\rho^{
		{(\kappa-1)\ell}/{\kappa}
		}\,.
\end{equation}
We seek next an upper bound on the value of~$\lambda$.
The Perron-Frobenius eigenvalue of the matrix
$N_\lambda$ is equal to~$1$, and all its entries are non-negative, thus 
	$$1\,\leq\,\max_{1\leq i\leq k}\sum_{1\leq j\leq k}N_{\lambda}(i,j)\,.$$
In particular, there exists an index $i\in\{\,1,\dots,k\,\}$ such that
\begin{equation}
\label{srinor}
	1\,\leq\,\sum_{1\leq j\leq k}N_{\lambda}(i,j)\,.
\end{equation}
Let us study the non-diagonal elements of the matrix $N_\lambda$. 
For $i\neq j$, we have
$$N_\lambda(i,j)\,=\,
	(\sigma_i-1)
    \sum_{n\geq 1} \frac{1}{\lambda^n}M^n(w^*_i,w^*_j)\,.
    $$
	In fact, the series depends only on the Hamming distance between 
	$w^*_i$ and $w^*_j$, as we can see from formula~\eqref{ucomput}.
	The same formula shows also that 
	$M^n(w^*_i,w^*_j)$ is a non-increasing function of $d(w^*_i,w^*_j)$.
	Therefore,
using the formula~\eqref{xcomput} with $d=1$,
	we have the inequality
\begin{equation}
    \label{bij}
N_\lambda(i,j)\,\leq\,
	(\sigma_i-1)
	\frac{\lambda(1-\rho)}{\kappa}
	\,
	\EE\Bigg(
	\displaystyle 
	\frac{ 1}{
	{\rho^{S_{\ell-1}}}
		\Big( \displaystyle \frac{\lambda}{\rho^{S_{\ell-1}}}-1\Big)
	\Big({ \displaystyle \frac{\lambda}{\rho^{S_{\ell-1}}}-\rho}
		\Big)}
		\Bigg)
	\,,  
\end{equation}
where $S_{\ell-1}$ is a binomial random variable with parameters $\ell-1,1-1/\kappa$. 
Moreover we know that $1\leq \lambda\leq\sigma$, thus
\begin{equation}
    \label{cij}
N_\lambda(i,j)\,\leq\,
	(\sigma_i-1)
	\sigma
	\frac{(1-\rho)}{\kappa}
	\,\EE\Bigg(
	\displaystyle 
	\frac{ 1}{ 
		\Big( \displaystyle \frac{\lambda}{\rho^{S_{\ell-1}}}-1\Big)
	\Big({ \displaystyle \lambda-{\rho^{S_{\ell-1}+1}}}
		\Big)}
		\Bigg)
	\,.  
\end{equation}
Plugging~\eqref{nij} and~\eqref{cij} in inequality~\eqref{srinor}, we get (recall that $\sigma\geq \sigma_i$)
\begin{equation}
    \label{cnj}
	\frac{1}{\sigma-1}
	\,\leq\,
	\EE\Bigg(
	\displaystyle 
	\displaystyle \frac{
		1}{
			\displaystyle \frac{\lambda}{\rho^{S_{\ell}}}-1}
		\Bigg)
	\, +\, 
	k\sigma \frac{(1-\rho)}{\kappa}
	\,\EE\Bigg(
	\displaystyle 
	\frac{ 1}{ 
		\Big( \displaystyle \frac{\lambda}{\rho^{S_{\ell-1}}}-1\Big)
	\Big({ \displaystyle \lambda-{\rho^{S_{\ell-1}+1}}}
		\Big)}
		\Bigg)
	\,.  
\end{equation}
We proceed in a way similar to what we did for the sharp peak landscape.
Namely, 
we pick a positive number $\varepsilon$ and
we split the expectations according to the values of $S_\ell$ and $S_{\ell-1}$, as follows:
\begin{multline}
\label{jnmbda}
	\frac{1}{\sigma-1}
	\,\leq\, 
	\frac{1}{ \displaystyle {\lambda}-1}
	\PP\Big(\Big|S_\ell-
	\frac{\kappa-1}{\kappa}\ell\Big|>\varepsilon\ell\Big)
	+
	\frac{1}{ \displaystyle \frac{\lambda}{
		\rho^{(\kappa-1)\ell/\kappa-\varepsilon\ell} }-1}\cr
	+
	 \frac{k\sigma(1-\rho)}{\kappa
	 (\displaystyle {\lambda}-1)
	 (\displaystyle {\lambda}-\rho)}
	\PP\Big(\Big|S_{\ell-1}-
	\frac{\kappa-1}{\kappa}(\ell-1)\Big|>\varepsilon(\ell-1)\Big)\cr
	+
	 \frac{k\sigma(1-\rho)}{\kappa}
	\frac{1}{ \Big(\displaystyle \frac{\lambda}{ \rho^{(\kappa-1)(\ell-1)/\kappa-\varepsilon\ell} }-1\Big)
	 \Big(\displaystyle {\lambda}-{ \rho^{(\kappa-1)(\ell-1)/\kappa-\varepsilon\ell} }\Big)}
\,.
\end{multline}
We use the estimate given by Chebyshev's inequality~\eqref{cheb} and we obtain
\begin{multline}
	\frac{1}{\sigma-1}
	\,\leq\, 
	\frac{1}{ \displaystyle {\lambda}-1}
	\frac{\kappa-1}{\kappa^2\varepsilon^2\ell}
	+
	\frac{1}{ \displaystyle \frac{\lambda}{
		\rho^{(\kappa-1)\ell/\kappa-\varepsilon\ell} }-1}
	+
	 \frac{k\sigma(1-\rho)}{\kappa
	 (\displaystyle {\lambda}-1)
	 (\displaystyle {\lambda}-\rho)}
	\frac{\kappa-1}{\kappa^2\varepsilon^2(\ell-1)}\cr
	+
	 \frac{k\sigma(1-\rho)}{\kappa}
	\frac{1}{ \Big(\displaystyle \frac{\lambda}{ \rho^{(\kappa-1)(\ell-1)/\kappa-\varepsilon\ell} }-1\Big)
	 \Big(\displaystyle {\lambda}-{ \rho^{(\kappa-1)(\ell-1)/\kappa-\varepsilon\ell} }\Big)}
\,.
\end{multline}
We regroup the first and the third term together, as well as the second and the fourth term and
we finally get (using that $1-\rho\leq \lambda-\rho$ and that $\rho\leq 1$)
\begin{multline}
\label{knmbda}
	\frac{1}{\sigma-1}
	\,\leq\, 
	 \frac{\kappa+k\sigma}{\kappa
	 \displaystyle ({\lambda}-1) }
	\frac{\kappa-1}{\kappa^2\varepsilon^2(\ell-1)}\cr
	+
	\frac{1}{ \displaystyle \frac{\lambda}{
		\rho^{(\kappa-1)(\ell-1)/\kappa-\varepsilon\ell} }-1}
	\Bigg(1+
	\frac{ \displaystyle\frac{k\sigma(1-\rho)}{\kappa} }{ 
	 \Big(\displaystyle {\lambda}-{ \rho^{(\kappa-1)(\ell-1)/\kappa-\varepsilon\ell} }\Big)}
	 \Bigg)
\,.
\end{multline}
By making an adequate choice for $\varepsilon$,
we shall now deduce an upper bound on $\lambda$ from this inequality.
\begin{lemma}
\label{extle}
There exists a positive constant $c$ such that, in the long chain regime,
the mean fitness $\lambda$ satisfies
\begin{equation}
\label{ymajor}
\lambda
\,\leq\,
	\max\Bigg(1+\frac{c}{\ell^{1/4}},
\sigma \Big(1-\displaystyle \frac{\kappa q}{\kappa-1}\Big)^{
	{(\kappa-1)(\ell-1)}/{\kappa}-\ell^{2/3}}\Big(1+\frac{1}{\ell^{1/13}}\Big)\Bigg)\,.
\end{equation}
\end{lemma}
\noindent
Before proving lemma~\ref{extle}, we explain the end of the
technical proof.
The lower bound~\eqref{mrinor} on $\lambda$ and 
the upper bound~\eqref{ymajor} on $\lambda$ yield the desired conclusion and this terminates the technical proof.
We have also obtained some estimates on the speed of convergence of $\lambda$, which could be improved with some
additional work.
\begin{proof}[Proof of lemma \ref{extle}] 
Suppose that $\lambda$ is larger than the upper bound in~\eqref{ymajor}. 
	We check that this is not compatible with inequality~\eqref{knmbda} with the choice $\varepsilon={\ell}^{-1/3}$. 
	Indeed, we have on one hand
	$$\lambda>
	1+\frac{c}{\ell^{1/4}}
	\qquad\Longrightarrow\qquad
	 \frac{\kappa+k\sigma}{\kappa
	 \displaystyle ({\lambda}-1) }
	\frac{\kappa-1}{\kappa^2 \ell^{-2/3} (\ell-1)}
	\,=\,O\Big( \frac{1}{\ell^{1/12}}\Big)\,.$$
		On the other hand,
\begin{multline*}
	\lambda>
\sigma \Big(1-\displaystyle \frac{\kappa q}{\kappa-1}\Big)^{
	{(\kappa-1)(\ell-1)}/{\kappa}-\ell^{2/3}}\Big(1+\frac{1}{\ell^{1/13}}\Big)
	\qquad\Longrightarrow\qquad\cr
	\exists c'>0\qquad
	\frac{1}{ \displaystyle \frac{\lambda}{
		\rho^{(\kappa-1)(\ell-1)/\kappa-\ell^{2/3}} }-1}
	\,\leq\,
	\frac{1}{ \displaystyle \sigma -1}
	-\frac{c'}{\ell^{1/13} }
	+o\Big( \frac{1}{\ell^{1/13} } \Big)
	\,.
\end{multline*}
We have also that (notice that we use here the fact that $a$ is finite)
\begin{equation*}
	\frac{ \displaystyle\frac{k\sigma(1-\rho)}{\kappa} }{ 
	 \Big(\displaystyle {\lambda}-{ \rho^{(\kappa-1)(\ell-1)/\kappa-\ell^{2/3}} }\Big)}
	 \,=\,
	O\Big( \frac{1}{\ell}\Big)\,.
\end{equation*}
	Combining the previous inequalities, we obtain that the right-hand member of inequality~\eqref{knmbda}
	has the following expansion:
	$$O\Big( \frac{1}{\ell^{1/12}}\Big)+
	\frac{1}{ \displaystyle \sigma -1}
	-\frac{c'}{\ell^{1/13} }
	+o\Big( \frac{1}{\ell^{1/13} } \Big)
+
	O\Big( \frac{1}{\ell}\Big)\,.
	$$
	This quantity becomes strictly less than $1/(\sigma-1)$ for $\ell$ large enough,
	and this is not compatible with
	the inequality~\eqref{knmbda}.
\end{proof}

\section{Plateau}
\label{late}
This section is devoted to the proof of theorems~\ref{plart}
and~\ref{palart}.
We first show that the value
$\lambda(a,\sigma)$ is well-defined. Then we perform a lumping procedure and we study the asymptotics of the reduced equations.
This yields the convergence 
of $\lambda$ towards $\lambda(a,\sigma)$.
The other claims of the theorem are proved with a similar argument.

\subsection{Definition of $\lambda(a,\sigma)$}
We start by proving that equation~\eqref{fiteq} admits a unique solution.
	Let us define the function $\phi_n$ by
\begin{equation}
	\label{hyper}
	\phi_n(a)\,=\,
	e^{-an}
	\sum_{k\geq 0}
	\frac{(an)^{2k}}{2^{2k}(k!)^2}\,.
\end{equation}
For any $a>0$, we have
$$ \phi_n(a)
	\,=\, e^{-an}
	\sum_{k\geq 0}
	\Bigg( \frac{1}{k!} \Big(
	\frac{an}{2}\Big)^k\Bigg)^2
	\,\leq\, e^{-an}
	\Bigg(\sum_{k\geq 0}
	 \frac{1}{k!} \Big(
	\frac{an}{2}\Big)^k\Bigg)^2\,=\,1
	\,.
	$$
It follows that the map
$$\lambda\mapsto
	\sum_{n\geq 1}
	\frac{{\phi_n(a)}}{\lambda^n}
	$$
	is continuous decreasing on $]1,+\infty[$. Moreover it converges to $0$ when $\lambda$ goes
	to $\infty$ and to $+\infty$ when $\lambda$ goes to $1$, because
	$$\displaylines{
		\sum_{n\geq 1}
	{{\phi_n(a)}}\,\geq\,
	\sum_{n\geq 1}
	e^{-an}
	\sum_{k\geq 0}
	\frac{(an)^{2k}}{(2k+1)!}\hfil\cr
	\hfil\,=\,
	\sum_{n\geq 1}
	\frac{e^{-an}}{an}
	\sum_{k\geq 0}
	\frac{(an)^{2k+1}}{(2k+1)!}
	\,=\,
	\sum_{n\geq 1}
	\frac{e^{-an}}{an}\text{sh}(an)\,=\,+\infty
	\,.}$$
Thus there exists a unique value $\lambda(a,\sigma)$ satisfying equation~\eqref{fiteq}.
We prove next the main claim of the theorem.
Throughout the proof, we write simply $\lambda(\ell,q)$ or even $\lambda$ instead of
	$\lambda_{\text{PL}}(\ell,q)$.
\subsection{Lumping}
We are dealing with a fitness function which depends only on the Hamming classes. So we lump together
the genotypes according to their Hamming classes and we introduce the new variables
$y(h)$, $0\leq h\leq\ell$, given by
$$\forall h\in\zl\qquad y(h)\,=\,\sum_{u:H(u)=h}x(u)\,.$$
Using formula~\eqref{eigendvpe}, we have
\begin{multline}
\label{ffvw}
	\forall h\in\zl\qquad y(h)
	\,=\, 
	\sum_{u:H(u)=h}
	\sum_{v} x(v)\big(f(v)-1\big)
    \sum_{n\geq 1} \frac{1}{\lambda^n}M^n(v,u)\,\cr
	\,=\,
	\sum_{v}
	x(v)\big(f(v)-1\big)
    \sum_{n\geq 1} \frac{1}{\lambda^n}
	\sum_{u:H(u)=h}
	M^n(v,u)\,.
\end{multline}
Now, the point is that the most inner sum depends only on the Hamming class of~$v$.
Indeed,
let us fix $b\in\zl$ and let $v$ be such that $H(v)=b$.
After $n$ mutation steps starting from $v$, we obtain a genotype $X_n$
whose components are $\ell$ independent Bernoulli
random variables. Among them, exactly $b$ 
have parameter $(1+\rho^n)/2$
and $\ell-b$
have parameter $(1-\rho^n)/2$ (this is a consequence of the 
computations performed in section~\ref{mutdyn}).
Therefore the Hamming class of $X_n$ is distributed
as the sum of two independent binomial random variables
with respective parameters
$(b,(1+\rho^n)/2)$ and
$(\ell-b,(1-\rho^n)/2)$. We conclude that
\begin{multline}
\label{mha}
\forall\, b,c\in\zl\,,\quad
\forall v\in
	\lbrace\,0,1\,\rbrace^\ell\,,\qquad\cr
	H(v)=b\quad\Longrightarrow\quad
	\sum_{u:H(u)=c}
	M^n(v,u)\,=\,
M^n_H(b,c)
	\,,
\end{multline}
where the matrix $M^n_H(b,c)$ is given by
\begin{equation}
	\label{mhbc}
M^n_H(b,c)\,=\,
	\PP\Bigg( \bino\Big(b, \frac{1+\rho^n}{2}\Big) 
	\,+\,
	 \bino\Big(\ell-b, \frac{1-\rho^n}{2}\Big)\,=\,c \Bigg)\,,
\end{equation}
$\bino(n,p)$ being a generic binomial random variable with parameters $(n,p)$
and the two binomial random variables appearing above being independent.
From the definition~\eqref{fplat}, we see that the fitness
$f_{\text{PL}}$ depends only on the Hamming, i.e.,
there exists a function $f$ such that
$f_{\text{PL}}(v)=f(H(v))$ for any $v$. 
With a slight abuse of notation, 
we still denote this function by $f_{\text{PL}}$.
Rearranging the sum in formula~\eqref{ffvw} according to the Hamming class of $v$
and using~\eqref{mha}, we get
\begin{align}
\label{rffvw}
	\forall h\in\zl\qquad 
	y(h)
	&\,=\,
	\sum_{0\leq k\leq\ell}
	\sum_{v:H(v)=k}
x(v)\big(
	f_{\text{PL}}
(v)-1\big)
    \sum_{n\geq 1} \frac{1}{\lambda^n}M_H^n(k,h)\,\cr
	&\,=\,
	\sum_{0\leq k\leq\ell}
	y(k)\big(
	f_{\text{PL}}
	(k)-1\big)
    \sum_{n\geq 1} \frac{1}{\lambda^n}
	M_H^n(k,h)\,.
\end{align}
Equation~\eqref{rffvw} is in fact valid as long as the fitness function depends only on the Hamming classes.
In the case of the plateau landscape, the equation for 
$h=\ell/2$ yields
\begin{equation}
	\label{fpla}
	\frac{1}{\sigma-1}
	\,=\,
    \sum_{n\geq 1} \frac{1}{\lambda^n}
	M^n_H\Big(\frac {\ell}{2},
	\frac {\ell}{2}\Big)\,.
\end{equation}
This is the counterpart of equation~\eqref{lambda} for the sharp peak landscape.
\subsection{Asymptotics of $M^n_H({\ell}/{2},\ell/2)$}
We shall rely on equation~\eqref{fpla} to analyze the asymptotic behavior of $\lambda$
in the long chain regime.
Let us begin with the term
	$M^n_H({\ell}/{2},\ell/2)$.
\begin{lemma}
	\label{mhco}
	In the long chain regime, we have the convergence
\begin{equation}
\label{mll}
\forall n\geq 1\qquad
\lim_{ \genfrac{}{}{0pt}{1}{\ell\to\infty,\, q\to 0 } {{\ell q} \to a } }
	M^n_H\Big(\frac {\ell}{2},
	\frac {\ell}{2}\Big)\,=\,
	\phi_n(a)\,,
\end{equation}
	where the sequence of functions $\phi_n$ is defined in~\eqref{hyper}.
\end{lemma}
\begin{proof}
	From formula~\eqref{mhbc} with $b=c=\ell/2$, we have
\begin{align*}
	\label{hbc}
	M^n_H\Big(
	\frac {\ell}{2},
	\frac {\ell}{2}\Big)
	&\,=\,
	\PP\Bigg( \bino\Big(
	\frac {\ell}{2},
	 \frac{1+\rho^n}{2}\Big) 
	\,+\,
	 \bino\Big(
	\frac {\ell}{2},
	  \frac{1-\rho^n}{2}\Big)\,=\,
	\frac {\ell}{2}
	 \Bigg)\cr
	&\,=\,
	\PP\Bigg( \bino\Big(
	\frac {\ell}{2},
	 \frac{1-\rho^n}{2}\Big) 
	\,=\,
	 \bino\Big(
	\frac {\ell}{2},
	  \frac{1-\rho^n}{2}\Big)
	 \Bigg)\,,
\end{align*}
with the understanding that the two binomial random variables appearing
above are independent.
Moreover, we have, for $a>0$,
	$$\frac {\ell}{2}
 \frac{1-\rho^n}{2}\,=\,
	\frac {\ell}{2}
	  \frac{1-(1-2q)^n}{2}\,\sim\,\frac{\ell}{2} nq\,\sim\,\frac{na}{2}\,,$$
hence we are in the regime where these binomial laws converge towards
the Poisson distribution $\cP(na/2)$ with parameter~$na/2$.
If $a=0$, the limit distribution is a Dirac mass at $0$.
Let us fix $K\geq 1$. We have
\begin{equation*}
	M^n_H\Big(
	\frac {\ell}{2},
	\frac {\ell}{2}\Big)
	\,\geq\,
	\sum_{0\leq k\leq K}
	\PP\Bigg( \bino\Big(
	\frac {\ell}{2},
	 \frac{1-\rho^n}{2}\Big) \,=\,k\Bigg)^2\,, 
\end{equation*}
whence, passing to the limit,
\begin{equation*}
\liminf_{ \genfrac{}{}{0pt}{1}{\ell\to\infty,\, q\to 0 } {{\ell q} \to a } }
	M^n_H\Big(\frac {\ell}{2},
	\frac {\ell}{2}\Big)\,\geq\,
	e^{-an}
	\sum_{0\leq k\leq K}
	\frac{(an)^{2k}}{2^{2k}(k!)^2}\,.
\end{equation*}
Sending $K$ to $\infty$, we obtain that
\begin{equation}
\label{limi}
\liminf_{ \genfrac{}{}{0pt}{1}{\ell\to\infty,\, q\to 0 } {{\ell q} \to a } }
	M^n_H\Big(\frac {\ell}{2},
	\frac {\ell}{2}\Big)\,\geq\,
	e^{-an}
	\sum_{k\geq 0}
	\frac{(an)^{2k}}{2^{2k}(k!)^2}\,=\,\phi_n(a)\,.
\end{equation}
Let us look for the reverse inequality. We fix again $K\geq 1$ and we write
\begin{equation*}
	M^n_H\Big(
	\frac {\ell}{2},
	\frac {\ell}{2}\Big)
	\,\leq\,
	\sum_{0\leq k\leq K}
	\PP\Bigg( \bino\Big(
	\frac {\ell}{2},
	 \frac{1-\rho^n}{2}\Big) \,=\,k\Bigg)^2\!\!+\, 
	\PP\Bigg( \bino\Big(
	\frac {\ell}{2},
	 \frac{1-\rho^n}{2}\Big) \,>\,K\Bigg)\,.
\end{equation*}
We bound the last term with the help of
Markov's inequality:
\begin{equation*}
	\PP\Bigg( \bino\Big(
	\frac {\ell}{2},
	 \frac{1-\rho^n}{2}\Big) \,>\,K\Bigg)\,\leq\,
	 \frac{1}{K} \frac {\ell}{2} \frac{1-\rho^n}{2}\,\leq\,
	 \frac{an}{K}
	 \,,
\end{equation*}
where the last inequality holds asymptotically.
From the previous two inequalities, we conclude that
\begin{equation*}
\limsup_{ \genfrac{}{}{0pt}{1}{\ell\to\infty,\, q\to 0 } {{\ell q} \to a } }
	M^n_H\Big(\frac {\ell}{2},
	\frac {\ell}{2}\Big)\,\leq\,
	e^{-an}
	\sum_{0\leq k\leq K}
	\frac{(an)^{2k}}{2^{2k}(k!)^2}\,+\,
	 \frac{an}{K}
	\,.
\end{equation*}
We finally send $K$ to $\infty$ to complete the proof of the lemma.
\end{proof}
\subsection{Convergence of $\lambda$}
	Let $\lambda^*$ be an accumulation point of $\lambda$.
	By definition,
there exist 
	two sequences of parameters $(\ell_m)_{m\geq 0}$,
	$(q_m)_{m\geq 0}$ such that
\begin{equation*}
\ell_m\to+\infty\,,\qquad
q_m\to 0\,,\qquad
\ell_m q_m\to a\,,\qquad
	\lambda(\ell_m, q_m)\to \lambda^*\,.
\end{equation*}
By lemma~\ref{mhco} and Fatou's lemma, we have
\begin{align}
	\sum_{n\geq 1} \frac{ \phi_n(a) }{(\lambda^*)^n}&\,=\,
    \sum_{n\geq 1} 
    \lim_{m\to\infty}
    \frac{1}{\lambda(\ell_m, q_m)^n}
	M^n_H\Big(\frac {\ell_m}{2},
	\frac {\ell_m}{2}\Big)\cr
	&\,\leq\,
    \lim_{m\to\infty}
    \sum_{n\geq 1} 
    \frac{1}{\lambda(\ell_m, q_m)^n}
	M^n_H\Big(\frac {\ell_m}{2},
	\frac {\ell_m}{2}\Big)\,=\,
    \frac{1}{\sigma-1}\,,
\end{align}
where the last equality comes from~\eqref{fpla}.
This inequality and the very definition of $\lambda(a,\sigma)$
(see equality~\eqref{fiteq})
imply that $\lambda^*\geq\lambda(a,\sigma)>1$.
In particular, there exist
$\varepsilon>0$ and $M\geq 1$ such that
\begin{equation}
\label{wfg}
	\forall m\geq M\qquad
	\lambda(\ell_m,q_m)\,\geq\,
1+\varepsilon\,.
\end{equation}
This last inequality
ensures that the series
    $$\sum_{n\geq 1} 
    \frac{1}{\lambda(\ell_m, q_m)^n}
	M^n_H\Big(\frac {\ell_m}{2},
	\frac {\ell_m}{2}\Big)$$
	converges uniformly with respect to $m$, so that we can interchange the limits and the infinite sum
	to get
\begin{align}
	\frac{1}{\sigma-1}&\,=\,
    \lim_{m\to\infty}
    \sum_{n\geq 1} 
    \frac{1}{\lambda(\ell_m, q_m)^n}
	M^n_H\Big(\frac {\ell_m}{2},
	\frac {\ell_m}{2}\Big)
	\cr
			  &\,=\,
    \sum_{n\geq 1} 
    \lim_{m\to\infty}
    \frac{1}{\lambda(\ell_m, q_m)^n}
	M^n_H\Big(\frac {\ell_m}{2},
	\frac {\ell_m}{2}\Big)
	\,=\,
    \sum_{n\geq 1} 
    \frac{
	\phi_n(a)
    }{(\lambda^*)^n}\,.
\end{align}
	This way we see that $\lambda^*$
	has to be the solution $\lambda(a,\sigma)$ of equation~\eqref{fiteq}. 
	In conclusion, the only possible accumulation point for $\lambda$
	in the long chain regime is $\lambda(a,\sigma)$. Therefore $\lambda$
	converges towards $\lambda(a,\sigma)$ as announced,
	and this completes the proof of theorem~\ref{plart}.

\subsection{Asymptotics of 
	$M^n_H\big(\lfloor {\alpha\ell}\rfloor, \lfloor {\alpha\ell}\rfloor \big)$}
The next lemma is the missing ingredient to complete the proof of theorem~\ref{palart}.
\begin{lemma}
	\label{smhco}
	In the long chain regime, we have the convergence
\begin{equation}
\label{smll}
\forall n\geq 1\qquad
\lim_{ \genfrac{}{}{0pt}{1}{\ell\to\infty,\, q\to 0 } {{\ell q} \to a } }
	M^n_H\big(\lfloor {\alpha\ell}\rfloor, \lfloor {\alpha\ell}\rfloor \big)
	\,=\,
	\phi_{n,\alpha}(a)\,,
\end{equation}
	where the sequence of functions $\phi_{n,\alpha}$ is defined by
\begin{equation}
	\label{ahyper}
	\phi_{n,\alpha}(a)\,=\,
	e^{-an}
	\sum_{k\geq 0}
	\frac{(\sqrt{4\alpha(1-\alpha)}an)^{2k}}{2^{2k}(k!)^2}
	\,.
\end{equation}
\end{lemma}
\begin{proof}
	From formula~\eqref{mhbc} with $b=c=\lfloor\alpha\ell\rfloor$, we have
\begin{align*}
	\label{hbc}
	M^n_H\big(\lfloor {\alpha\ell}\rfloor, \lfloor {\alpha\ell}\rfloor \big)
	&\,=\,
	\PP\Big( \bino\Big(
	\lfloor {\alpha\ell}\big\rfloor,
	 \frac{1+\rho^n}{2}\Big) 
	\,+\,
	\bino\Big(
	{\ell}-
	\lfloor {\alpha\ell}\rfloor,
	  \frac{1-\rho^n}{2}\Big)\,=\,
	\lfloor {\alpha\ell}\rfloor
	 \Big)\cr
	&\,=\,
	\PP\Big( 
	 \bino\Big(
	{\ell}-
	\lfloor {\alpha\ell}\rfloor,
	  \frac{1-\rho^n}{2}\Big)\,=\,
	 \bino\Big(
	\lfloor {\alpha\ell}\rfloor,
	  \frac{1-\rho^n}{2}\Big)
	 \Big)\,,
\end{align*}
with the understanding that the two binomial random variables appearing
above are independent.
Moreover, we have
	$${\ell}
 \frac{1-\rho^n}{2}\,=\,
	{\ell}
	  \frac{1-(1-2q)^n}{2}\,\sim\,{\ell} nq\,\sim\,{na}\,,$$
hence we are in the regime where the first binomial law converges towards
	the Poisson distribution with parameter $(1-\alpha)na$ while the 
second binomial law converges towards
	the Poisson distribution with parameter $\alpha na$.
	We proceed as in the proof of lemma~\ref{mhco} to conclude that
\begin{equation*}
\forall n\geq 1\qquad
\lim_{ \genfrac{}{}{0pt}{1}{\ell\to\infty,\, q\to 0 } {{\ell q} \to a } }
	M^n_H\big(\lfloor {\alpha\ell}\rfloor, \lfloor {\alpha\ell}\rfloor \big)
	\,=\,
	e^{-an}
	\sum_{k\geq 0}
	\frac{((1-\alpha)an)^{k}}{k!}
	\frac{(\alpha an)^{k}}{k!}
	\,,
\end{equation*}
	and we rewrite the right-hand quantity as in formula~\eqref{ahyper}.
\end{proof}
\subsection{Completion of the proof of theorem~\ref{palart}}
Let $\phi_{\alpha}(a)$ be the function defined by
\begin{equation}\label{phial}
	\phi_{\alpha}(a)\,=\,
	\sum_{n\geq 1}
	\phi_{n,\alpha}(a)\,=\,
	\sum_{n\geq 1}
	e^{-an}
	\sum_{k\geq 0}
	\frac{(\sqrt{4\alpha(1-\alpha)}an)^{2k}}{2^{2k}(k!)^2}\,.\end{equation}
In order to study the behavior of $\phi_\alpha(a)$ as $a$ goes to $0$
or $\infty$, we take advantage of the simple inequalities
	$(2k)!\leq {2^{2k}(k!)^2}\leq (2k+1)!$
to bound 
	$\phi_{n,\alpha}(a)$ from above and from below as follows:
	$$
	e^{-an}
	\frac{
	\sinh\big(\sqrt{4\alpha(1-\alpha)}an\big)
	}{ \sqrt{4\alpha(1-\alpha)}an }
	\,\leq\,
	\phi_{n,\alpha}(a)\,\leq\,
	e^{-an}\cosh\big( \sqrt{4\alpha(1-\alpha)}an\big)\,.$$
	From these inequalities, we deduce that, when $\alpha\neq 1/2$, we have
\begin{equation}
	\label{limbd}
	\lim_{a\to 0} \phi_\alpha(a)\,=\,+\infty\,,\qquad
	\lim_{a\to +\infty} \phi_\alpha(a)\,=\,0\,.\
\end{equation}
Moreover, for any $a_0>0$, we have
\begin{equation*}
	\forall  a\geq a_0\qquad
	\phi_{n,\alpha}(a)\,\leq\,
	\exp\big({-a_0n}(1-\sqrt{4\alpha(1-\alpha)})\big)\,.
\end{equation*}
Thus, when $\alpha\neq 1/2$, 
the series
	$\sum_{n\geq 1}
	\phi_{n,\alpha}(a)$ is uniformly convergent over $[a_0,+\infty[$
	for any $a_0>0$
	and therefore the function $\phi_\alpha$ is continuous on $]0,+\infty[$.
	We consider the equation
	$$
	\frac{1}{\sigma-1}
	\,=\,\phi_\alpha(a)\,,$$
	and we denote by
	$a^1_c(\sigma,\alpha)$ (respectively 
	$a^2_c(\sigma,\alpha)$)
	the smallest
	(respectively the largest) solution to this equation in $]0,+\infty[$.
These two real numbers are well-defined, thanks to the limits~\eqref{limbd} and
the continuity of $\phi_\alpha$ on
$]0,+\infty[$. 

We use a strategy similar to the one of the proof of theorem~\ref{plart}.
Suppose that $a\geq a_c^2(\sigma,\alpha)$.
The equation for 
$h=\lfloor\alpha\ell\rfloor$ in the system~\eqref{rffvw}
	yields
\begin{equation}
	\label{fspla}
	\frac{1}{\sigma-1}
	\,=\,
    \sum_{n\geq 1} \frac{1}{\lambda^n}
	M^n_H\big(\lfloor {\alpha\ell}\rfloor, \lfloor {\alpha\ell}\rfloor \big)\,.
\end{equation}
Let $\lambda^*$ be an accumulation point of $\lambda$.
If $\lambda^*>1$, then, proceeding as in the proof of theorem~\ref{plart},
we pass to the limit along a subsequence in equation~\eqref{fspla} to get
\begin{equation}
	\label{lrfspla}
	\frac{1}{\sigma-1}
	\,=\,
	\sum_{n\geq 1} \frac{1}{(\lambda^*)^n}
	\phi_{n,\alpha}(a)\,<\,\phi_\alpha(a)
	\,,
\end{equation}
but this contradicts the fact that $a\geq a^2_c(\sigma,\alpha)$.
Suppose that $a< a_c^1(\sigma,\alpha)$.
Let $\lambda^*$ be an accumulation point of $\lambda$.
Suppose that $\lambda^*=1$.
By Fatou's lemma, we would have
\begin{equation}
	\label{flrfspla}
	\phi_\alpha(a)\,=\,
	\sum_{n\geq 1} 
	\phi_{n,\alpha}(a)\,
	\,\leq\,
\liminf_{ \genfrac{}{}{0pt}{1}{\ell\to\infty,\, q\to 0 } {{\ell q} \to a } }
    \sum_{n\geq 1} \frac{1}{\lambda^n}
	M^n_H\big(\lfloor {\alpha\ell}\rfloor, \lfloor {\alpha\ell}\rfloor \big)
	\,=\,
	\frac{1}{\sigma-1}
	\,,
\end{equation}
where the last equality comes from~\eqref{fspla}.
Yet inequality~\eqref{flrfspla}
stands in 
contradiction with the fact that $a<a^1_c(\sigma,\alpha)$.
%
Therefore the infimum limit of $\lambda$ has to be strictly larger than~$1$.
\section{Survival of the flattest}
\label{sof}
This final section is devoted to the proof of theorem~\ref{isurt}.
\subsection{Lower bound on
$\lambda_{\text{SP/PL}}(\delta,\sigma,\ell,q)$}
The mean fitness
	$\lambda_{\text{SP/PL}}(\delta,\sigma,\ell,q)$
	is also the Perron-Frobenius eigenvalue of
	the matrix $\smash{\big(
	f_{\text{SP/PL}}(u)
M(u,v), {u,v\in E} \big)}$. 
	As such, it is a non-decreasing function of the entries
	of this matrix. Therefore
	$$\lambda_{\text{SP/PL}}(\delta,\sigma,\ell,q)
	\,\geq\,\max\big(
	\lambda_{\text{SP/PL}}(\delta,1,\ell,q),
	\lambda_{\text{SP/PL}}(1,\sigma,\ell,q)\big)
	\,.$$
	With the 
	help of theorems~\ref{ifrrt} and~\ref{plart},
	we conclude that
	$$\liminf_{ \genfrac{}{}{0pt}{1}{\ell\to\infty,\, q\to 0 } {{\ell q} \to a } }
	\lambda_{\text{SP/PL}}(\delta,\sigma,\ell,q)
	\,\geq\,\max\big(\delta\exa,\lambda(a,\sigma)\big)\,.$$
Let us prove next the reverse inequality.
\subsection{Lumping}
Let $(x(u), u\in \lbrace\,0,1\,\rbrace^\ell)$ be the solution to the quasispecies equations
associated with $f_{\text{SP/PL}}$.
We are dealing with a fitness function which depends only on the Hamming classes. So we lump together
the genotypes according to their Hamming classes and 
we introduce the new variables
$y(h)$, $0\leq h\leq\ell$, given by
$$\forall h\in\zl\qquad y(h)\,=\,\sum_{u:H(u)=h}x(u)\,.$$
We use the same 
computations that lead from~\eqref{ffvw} to~\eqref{rffvw} 
(these computations relied on the fact that the fitness
function depended only on the Hamming classes)
to conclude
that 
the variables
$y(h)$, $0\leq h\leq\ell$, satisfy
\begin{equation}
\label{uffvw}
	\forall h\in\zl\qquad 
	y(h)
	\,=\,
	\sum_{0\leq k\leq\ell}
	y(k)\big(
	f_{\text{SP/PL}}
	(k)-1\big)
    \sum_{n\geq 1} \frac{1}{\lambda^n}
	M_H^n(k,h)\,,
\end{equation}
where the matrix $M_H(k,h)$ was defined in~\eqref{mhbc}.
Again,
with a slight abuse of notation, 
we denote by
$f_{\text{SP/PL}}(k)$ the common value of the fitness function
$f_{\text{SP/PL}}$ for all the sequences in the Hamming class $H(k)$.
In the sequel of the proof, we write simply $\lambda(\ell,q)$ or even $\lambda$ instead of
	$\lambda_{\text{SP/PL}}(\delta,\sigma,\ell,q)$.
The value $\lambda$ and the variables $y$ satisfy also
\begin{equation}
\label{lzlu}
	\lambda\,=\,\delta y(0)+\sigma y(\ell/2)+1-y(0)-y(\ell/2)
\end{equation}
and 
	$$\sum_{0\leq k\leq\ell}y(k)\,=\,1\,.$$
\subsection{The two main equations}
	Let us rewrite
	the two equations of the system~\eqref{uffvw} corresponding to
	$h=0$ and $h=\ell/2$:
\begin{align}
\label{zwvw}
	y({0}) &\,=\,  
	y(0)\big(\delta-1\big)
    \sum_{n\geq 1} \frac{1}{\lambda^n}
	M_H^n (0,0)
	\,+\,
	y\Big(\frac{\ell}{2}\Big) 
	\big(\sigma-1\big)
    \sum_{n\geq 1} \frac{1}{\lambda^n}
	M_H^n \Big(\frac{\ell}{2},{0}\Big) 
	\,, \cr
	y\Big(\frac{\ell}{2}\Big) &\,=\,  
	y(0)\big(\delta-1\big)
    \sum_{n\geq 1} \frac{1}{\lambda^n}
	M_H^n \Big({0},\frac{\ell}{2}\Big) 
	\,+\,
	y\Big(\frac{\ell}{2}\Big) 
	\big(\sigma-1\big)
    \sum_{n\geq 1} \frac{1}{\lambda^n}
	M_H^n \Big(\frac{\ell}{2},\frac{\ell}{2}\Big) 
	\,. 
\end{align}
	Let us examine what are the possible limits of $\lambda$, $y(0)$ and $y(\ell/2)$
	along a subsequence in the long chain regime. So, let us suppose that,
	along a subsequence, the following convergences take place:
\begin{equation}
	\label{gg}
	\lambda\to\lambda^*\,,\quad
	y(0)\to y_0^*\,,\quad
	y(\ell/2)\to y_1^*\,.
\end{equation}
	In the sequel of the argument, all the limits are taken along this subsequence.
	Passing to the limit in ~\eqref{lzlu}, we get
\begin{equation}
\label{llzlu}
	\lambda^*\,=\,\delta y_0^*+\sigma y_1^*+1-y_0^*-y_1^*\,.
\end{equation}
Since $\lambda^*\geq\lambda(a,\sigma)>1$, then necessarily
$y_0^*+y_1^*>0$.
Moreover, the following convergences take place:
\begin{equation}
\label{limd}
	M^n_H(0,0)
	\to e^{-na}\,,\quad
	M^n_H\Big(\frac {\ell}{2},
	\frac {\ell}{2}\Big)\to \phi_n(a)\,,\quad
\end{equation}
	while
\begin{equation}
\label{limz}
	M^n_H\Big(\frac {\ell}{2}, {0}\Big)\to 0\,,\quad
	M^n_H\Big({0}, \frac {\ell}{2}\Big)\to 0\,.
\end{equation}
The convergences~\eqref{limd} have already been proved. 
The first one is obvious and the second one
is the purpose of lemma~\ref{mhco}.
To prove the
convergences~\eqref{limz}, we use the expression of $M_H$ computed in~\eqref{mhbc} and we get
\begin{align*}
	M^n_H\Big(\frac {\ell}{2}, {0}\Big)
	&\,=\,
	\PP\Bigg( \bino\Big(\frac{\ell}{2}, \frac{1+\rho^n}{2}\Big) 
	 +\bino\Big(\frac{\ell}{2}, \frac{1-\rho^n}{2}\Big)
	\,=\,0\Bigg)\cr
	&\,\leq\,
	\PP\Bigg( \bino\Big(\frac{\ell}{2}, \frac{1+\rho^n}{2}\Big)\,=\,0 \Bigg)
	\,\leq\,
	\PP\Bigg( \bino\Big(\frac{\ell}{2}, \frac{1}{2}\Big) \,=\,0\Bigg)\,,
	\cr
	M^n_H\Big({0}, \frac {\ell}{2}\Big)
	&\,=\,
	 \PP\bigg(\bino\Big(\ell, \frac{1-\rho^n}{2}\Big)\,=\,\frac{\ell}{2} \Bigg)
	\,\leq\,
	 \PP\bigg(\bino\Big(\ell, \frac{1}{2}\Big)\,=\,\frac{\ell}{2} \Bigg)
	 \,,
\end{align*}
and it is well-known that the right-hand quantities goes to $0$ as $\ell$ goes to $\infty$.
\subsection{Upper bound on
$\lambda_{\text{SP/PL}}(\delta,\sigma,\ell,q)$}
	Proceeding as in the proof of theorem~\ref{plart} (see lemma~\ref{mhco} and thereafter), 
	we take advantage of the fact that $\lambda^*>1$ to interchange the limit and the four
	sums appearing in~\eqref{zwvw}. So, passing to the limit 
	in~\eqref{zwvw}, and using the convergences~\eqref{gg}, \eqref{limd}, \eqref{limz},
	we get
\begin{align}
\label{gwvw}
	y_0^* &\,=\,  
	y_0^*\big(\delta-1\big)
	\sum_{n\geq 1} \frac{e^{-na}}{(\lambda^*)^n}
	\,=\,
	y_0^*
	\frac{\delta-1}
	{{\lambda^*}{e^{a}}-1}
	\,, \cr
	y_1^*
	&\,=\,  
	y_1^*
	\big(\sigma-1\big)
	\sum_{n\geq 1} \frac{\phi_n(a)}{(\lambda^*)^n}
	\,. 
\end{align}
	If $y_0^*>0$, then the first equation implies that
	$\lambda^*=\delta \exa$.
If $y_1^*>0$, then the second equation is equivalent to
	equation~\eqref{fiteq}. Therefore $\lambda^*$ has to be equal to the unique solution
	$\lambda(a,\sigma)$ of this equation.
%
	In conclusion, there are only two possible limits along a subsequence for $\lambda$,
	namely
	$\delta\exa$ and $\lambda(a,\sigma)$, thus
	$$\limsup_{ \genfrac{}{}{0pt}{1}{\ell\to\infty,\, q\to 0 } {{\ell q} \to a } }
	\lambda_{\text{SP/PL}}(\delta,\sigma,\ell,q)
	\,\leq\,\max\big(\delta\exa,\lambda(a,\sigma)\big)\,.$$
	The final conclusions of the theorem are obtained as a by-product of the previous
	argument. 
	Indeed, if $\delta\exa\neq\lambda(a,\sigma)$, then it follows from~\eqref{gwvw}
	that we cannot have simultaneously $y_0^*>0$ and $y_1^*>0$.
	This remark, in conjunction with the identity~\eqref{llzlu}, yields the convergences
	stated at the end of the theorem.

\section*{Acknowledgements}
The authors are grateful to the Referees for their careful reading
and their numerous constructive comments:
not only did they detect several mathematical imprecisions, 
they also really helped us to improve the quality of the paper and
its overall presentation.

\bibliographystyle{plain}
\bibliography{quas}
 \thispagestyle{empty}

\end{document}